\documentclass[a4paper,fleqn]{cas-dc}
\usepackage[square,sort,comma,numbers]{natbib}
\usepackage{multirow}
\usepackage{hyperref}
\usepackage{graphicx,subcaption}
\usepackage{bookmark}
\usepackage{float}
\usepackage{orcidlink}

\definecolor{mcolor}{RGB}{34, 150, 209}
\hypersetup{colorlinks, allcolors=mcolor}

\begin{document}
\let\printorcid\relax
\let\WriteBookmarks\relax
\def\floatpagepagefraction{1}
\def\textpagefraction{.001}

\shorttitle{Classification of Breast Tumours Using Gradient Boosting Methods}    
\shortauthors{Abbasniya et al.}  
\title[mode = title]{Classification of Breast Tumours Based on Histopathology Images Using Deep Features and Ensemble of Gradient Boosting Methods}

\author[1]{Mohammad Reza Abbasniya$^{\orcidlink{0000-0003-0806-2858}}~$}[orcid=0000-0003-0806-2858]
\fnmark[1] 

\author[1]{Sayed Ali Sheikholeslamzadeh$^{\orcidlink{0000-0001-6898-813X}}~$}[orcid=0000-0001-6898-813X]
\fnmark[1] 

\author[2]{Hamid Nasiri$^{\orcidlink{0000-0002-9279-6063}}~$}[orcid=0000-0002-9279-6063]
\cormark[1]

\address[1]{{Department of Electrical and Computer Engineering, Semnan University}, Semnan, Iran}

\address[2]{{Department of Computer Engineering, Amirkabir University of Technology}, Tehran, Iran}

\author[1]{Samaneh Emami$^{\orcidlink{0000-0002-3879-7188}}~$}[orcid=0000-0002-3879-7188]

\cortext[cor1]{Correspondence should be addressed to Hamid Nasiri; \href{mailto:h.nasiri@aut.ac.ir}{h.nasiri@aut.ac.ir}}
\fntext[fn1]{These authors contributed equally to this work and should be regarded as co-first authors.}

\begin{abstract}[S U M M A R Y]
	
Breast cancer is the most common cancer among women worldwide. Early-stage diagnosis of breast cancer can significantly improve the efficiency of treatment. Computer-aided diagnosis (CAD) systems are widely adopted in this issue due to their reliability, accuracy and affordability. There are different imaging techniques for a breast cancer diagnosis; one of the most accurate ones is histopathology which is used in this paper.
Deep feature transfer learning is used as the main idea of the proposed CAD system’s feature extractor. Although 16 different pre-trained networks have been tested in this study, our main focus is on the classification phase. The Inception-ResNet-v2 which has both residual and inception networks profits together has shown the best feature extraction capability in the case of breast cancer histopathology images among all tested CNNs. In the classification phase, the ensemble of CatBoost, XGBoost and LightGBM has provided the best average accuracy.
The BreakHis dataset was used to evaluate the proposed method. BreakHis contains 7909 histopathology images (2,480 benign and 5,429 malignant) in four magnification factors. The proposed method’s accuracy (IRv2-CXL) using 70\% of BreakHis dataset as training data in 40×, 100×, 200× and 400× magnification is 96.82\%, 95.84\%, 97.01\% and 96.15\%, respectively.
Most studies on automated breast cancer detection have focused on feature extraction, which made us attend to the classification phase. IRv2-CXL has shown better or comparable results in all magnifications due to using the soft voting ensemble method which could combine the advantages of CatBoost, XGBoost and LightGBM together.

\end{abstract}

\begin{keywords} 
	BreakHis \sep 
	Breast Cancer \sep 
	Ensemble classification \sep 
	Grad-CAM  \sep 
	Inception-ResNet-v2 \sep 
	Transfer learning
\end{keywords}

\maketitle

\section{Introduction}
The term cancer is a generic word for an extensive group of diseases which may affect different parts of the body. To understand cancer, we must first know about Tumours. Tumours can either be benign or malignant which are recognized as cancerous. Tumours that are deemed to be benign (noncancerous) increase in size slowly and do not spread. Cancerous tumours grow swiftly, overrun and dismantle nearby healthy tissues throughout the whole body \cite{abdul2020biomaterials}. Breast cancer arises in lining cells of ducts or lobules in the glandular tissue of the breast \cite{harmer2008breast}. According to the WHO, breast cancer is the most prevalent cancer worldwide and the fifth most common cause of cancer death in 2020
\footnote{\url{https://www.who.int/news-room/fact-sheets/detail/cancer}, Retrieved on 2022-2-20}. If breast cancer is detected at its early stages, it can be treated quite effectively before it spreads to other parts of the body \cite{hajiabadi2020combination}. \par
Breast cancer is diagnosed using various imaging techniques such as mammography, ultrasound, thermography and pathological tests \cite{mohammed2018neural, shahidi2020breast}. Among these methods, for those patients who have undergone other forms of scannings, histopathology images are considered to be the gold standard for improving the accuracy of results as well as providing reliable information for assessing the effects of cancer on surrounding tissue \cite{shahidi2020breast}. Histopathological images of tissues from breast tumour patients are obtained by applying certain chemicals to dye the nucleus of the cells and then dyeing the other components with another chemical in different shades to accentuate various parts tissue structures and cellular characteristics \cite{shahidi2020breast}. \par
After completing the biopsy, diagnosis will be done by a pathologist who will examine the stained tissue using a microscope. Although these images are very comprehensive, due to various difficulties such as scant contrast in images, noise and lack of appreciation by the human eye the occurrence of misdiagnoses are fairly possible \cite{alanazi2021boosting}. Add to this a lack of well-trained specialists and the extremely time-consuming analysis of H\&E-stained images using microscopes and it becomes understandable why an alternative method to using human specialists is needed \cite{punitha2021automated}. With the onset of pattern recognition, machine learning (ML) and convolutional neural networks, to overcome existing problems, researchers are focusing on using CAD to improve tests accuracy \cite{shahidi2020breast}. CAD systems are classified into two different categories. One category sorts the systems by the staining method of images used as input, which are H\&E-stained and IHC images. Images obtained by using haematoxylin and eosin are more suited for determining whether a neoplasm is cancerous or not but analysis of these images calls for more sophisticated image processing and machine learning methods. Immunohistochemistry images which are marked by certain biomarkers to illustrate certain cells or regions are especially handy when cancer has already been diagnosed using the H\&E method. \cite{alirezazadeh2018representation}.\par
Another method for categorizing CAD systems is grouping them by whether the images they use are WSI or ROI. Both H\&E-stained and IHC images are used as input in these systems. For diagnosis using WSI as input, the entirety of images is used without any tampering. In systems that use ROI however key parts are concentrated on using segmentation or image detection \cite{alirezazadeh2018representation}. In analysing WSI’s, no segmentation or path determination are done on images with the goal of the system recognizing cancerous and noncancerous slides. With this fact in mind experts are only required to label the images in the training phase as either benign or malignant \cite{alirezazadeh2018representation}. CAD systems use different algorithms and structures for tumour classification; this study aims to propose a new method for binary classification of tumours based on histopathology whole slide images using deep feature transfer learning. \par
Transfer learning is a deep learning method where one models knowledge is passed on to another model \cite{perlich2014machine}. Using transfer learning algorithms reduces training time of models drastically, allowing for different solutions to be built straight away. Furthermore, it can save us from establishing an expensive and complicated cloud GPU/TPU. One other noteworthy benefit of using transfer learning reveals itself in tackling a lack of data. 
For deep learning models to solve a task successfully, a good deal of data is needed. However, cases, where data is abundantly available, are few and far between, leaving us to deal with a lack of data most of the time, as evident in the case of breast tumour classification \cite{lv2022exploratory, ahuja2021deep}. \par

\section{Related Works}
An early-stage breast cancer diagnosis can substantially decrease its mortality rate; as such, CAD systems using ML approaches could be widely effective here. As any other ML problem, researchers and data scientists need enough data to be involved in this context. Therefore, many different datasets have been provided with various imaging modalities such as Mammograms (MGs), Ultrasound (US), Magnetic Resonance Imaging (MRI) and Histopathology (HP) \cite{shah2022artificial}. \par
For example, the Bioimaging dataset containing 156 annotated H\&E-stained images in size 2048×1536 divided into four classes (normal, benign, in situ and invasive) is used in a study by Fondon et al. (2018) \cite{fondon2018automatic}. In their work, contrast level adaptive histogram equalization is used for preprocessing images and in the next step 250 numerical features based on nuclei, colour region and texture characteristics are extracted, and finally, an SVM classifier is adopted to classify images according to extracted features. \par
In another study by George et al. (2020) \cite{george2020breast}, 200× magnification images of the BreakHis \cite{spanhol2015dataset} dataset was used to evaluate their method. The NucTraL+BCF is their suggested method which consists of four sections; first the Macenko \cite{macenko2009method} strategy is used for stain normalization as the preprocessing phase, then nucleus paths are extracted from images and enter into three individual transfer learning-based processes. Each process utilizes an SVM classifier and in the final section, belief theoretical classifier fusion is used to determine the final result according to results provided by three SVM classifiers. \par
In addition, invasive ductal carcinoma (IDC) is an imbalanced dataset that contains 277,524 colour images divided into two classes (negative and positive). This dataset is used by Choudhary et al. (2021) \cite{choudhary2021transfer}. In their study, several CNNs and the pruned version of them were employed to train on 70\% of the dataset’s images by two strategies, training the whole CNN or only the last layer. The pruned version of ResNet50, in which all layers were trained, has shown the best performance in that case. Another study by Barsha et al. (2021) \cite{barsha2021automated} tried to use an ensemble of two pre-trained CNNs on the same dataset (IDC), but the training size was 80\% of the dataset. In that paper, a dense layer with softmax activation was added to the end of the DenseNet-121 and DenseNet-169 networks for binary classification. Test time augmentation was employed to increase model accuracy and the final decision was reached by using the mean of prediction probabilities. \par
Thermography infrared is another technique for medical imaging which can be used in the breast cancer diagnosis process. This technique is used in collecting DMR-IR dataset which is adopted widely to evaluate state-of-the-art breast cancer diagnosis methods such as the proposed methods by Gonçalves et al. (2022) \cite{gonccalves2022cnn} and Chatterjee et al. (2022) \cite{chatterjee2022breast}. Similar to other mentioned works, Gonçalves et al. also used pre-trained CNNs. They utilized original weighted networks of VGG-16, ResNet-50 and DenseNet-201 and then tried to search for the best architecture of the fully connected layer to be adopted as a binary classifier at the end of each CNN. In that paper, two bio-inspired algorithms (i.e., Genetic Algorithm and Particle swarm optimization) were suggested as searching methods. In the study by Chatterjee et al. \cite{chatterjee2022breast} pre-trained version of VGG-16 was used to extract image features as well. The distinctive point of their study is to use a memory-optimized version of the Dragonfly Algorithm to reduce the dimension of the features vector by about 40\% before adopting an SVM classifier. \par

\section{Materials and Methodology}
\subsection{Dataset}
\begin{figure*}[!h]
	\centering
	\begin{subfigure}[t]{0.22\textwidth} \includegraphics[height=25mm]{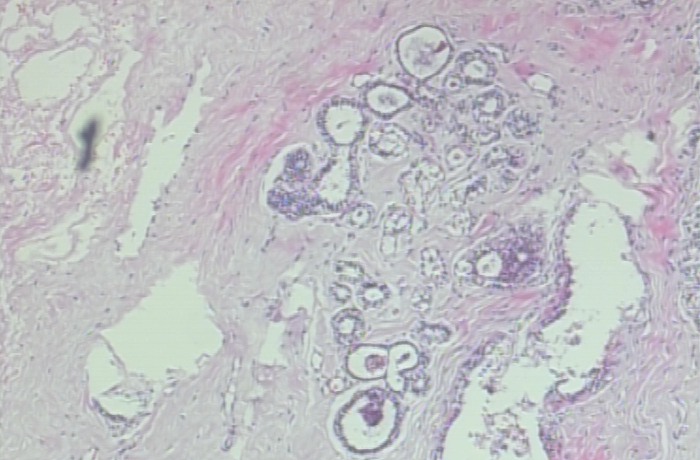} \caption{Adenosis}	      \end{subfigure}  \hfill
	\begin{subfigure}[t]{0.22\textwidth} \includegraphics[height=25mm]{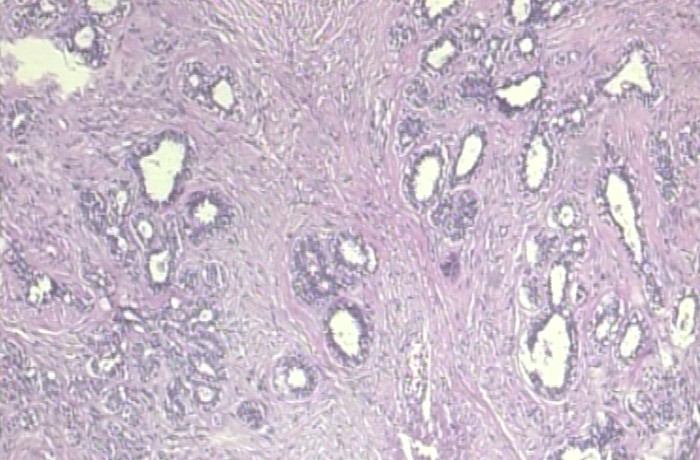} \caption{Fibroadenoma}	  \end{subfigure}  \hfill
	\begin{subfigure}[t]{0.22\textwidth} \includegraphics[height=25mm]{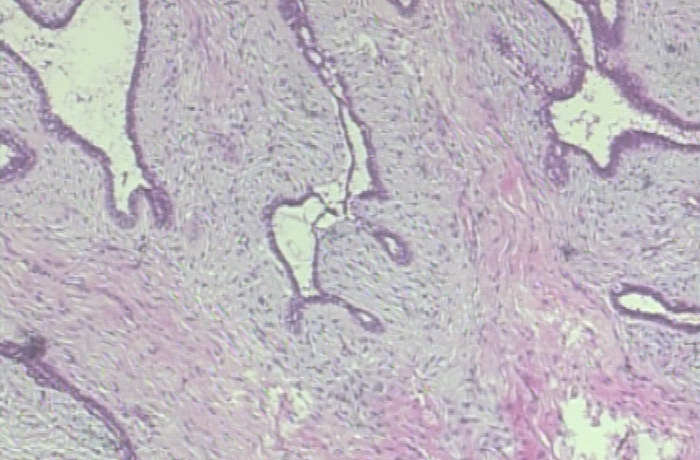} \caption{Tubular Adenoma} \end{subfigure}  \hfill
	\begin{subfigure}[t]{0.22\textwidth} \includegraphics[height=25mm]{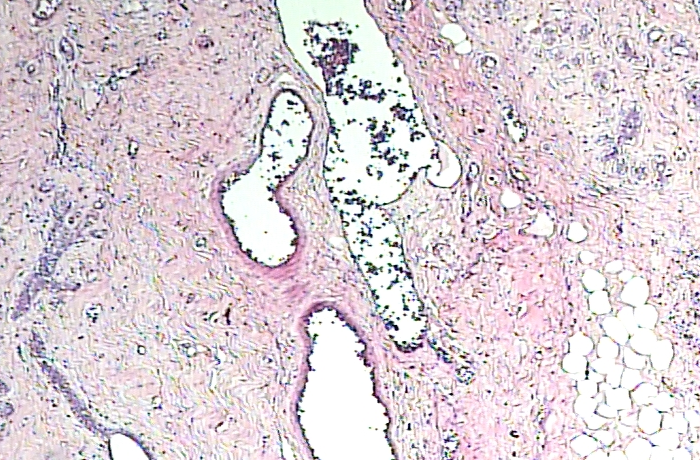} \caption{Phyllodes Tumour}\end{subfigure}  

	\par\bigskip

	\begin{subfigure}[t]{0.22\textwidth} \includegraphics[height=25mm]{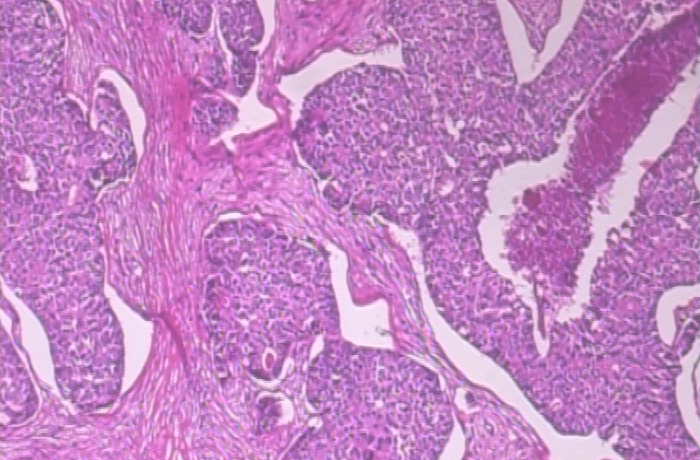} \caption{Ductal Carcinoma}	\end{subfigure}  \hfill
	\begin{subfigure}[t]{0.22\textwidth} \includegraphics[height=25mm]{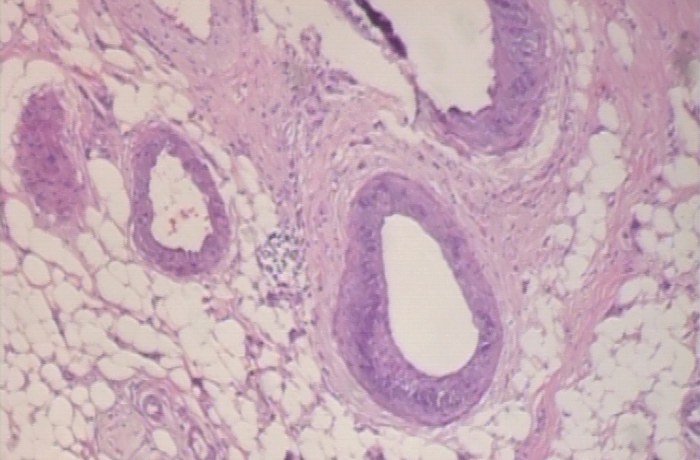} \caption{Lobular Carcinoma}	\end{subfigure}  \hfill
	\begin{subfigure}[t]{0.22\textwidth} \includegraphics[height=25mm]{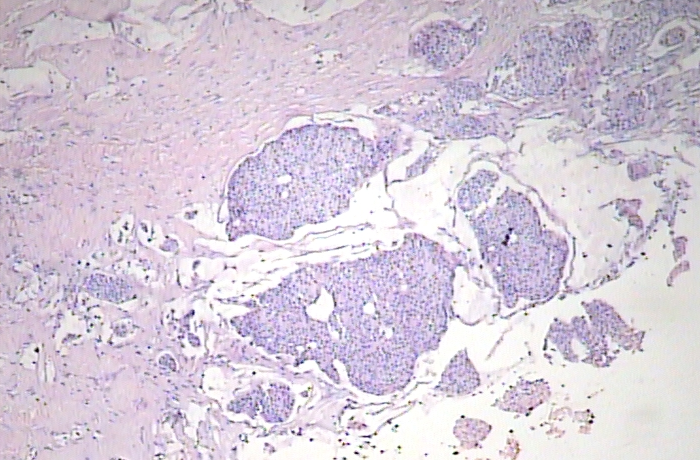} \caption{Mucinous Carcinoma}\end{subfigure}  \hfill
	\begin{subfigure}[t]{0.22\textwidth} \includegraphics[height=25mm]{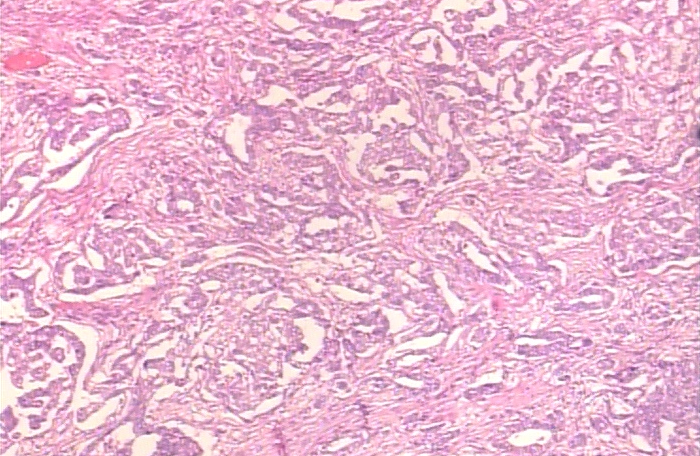} \caption{Papillary Carcinoma}\end{subfigure}  

	\caption{Samples of benign (a-d) and malignant (e-h) tumour types in 40× magnification.}
	\label{table2}
\end{figure*}

In this article Breast Cancer Histopathological Image Classification (BreakHis) dataset is used in order to examine the viability of our method \cite{spanhol2015dataset}. The BreakHis dataset was collected in a clinical study by Pathological Anatomy and Cytopathology Laboratory, Parana state, Brazil from January to December of 2014. BreakHis is a public and well-known dataset that contains 7,909 high-quality images taken from the microscopic biopsy of benign and malignant (2,480 and 5,429 images respectively) breast tumours. BreakHis dataset contains two procedures of biopsy: Surgical Open Biopsy (SOB) and Core Needle Biopsy (CNB) section of $\sim$ 3µm thickness. In addition, haematoxylin and eosin (HE) stained breast tissue biopsy slides are used in the process of collecting images. Image acquisition is done by Olympus BX-50 microscope with a 3.3× magnification relay lens coupled to SCC-131AN which is a Samsung digital colour camera with 6.5 µm pixel size. These images were collected from 82 patiens at four magnifying factors: 40×, 100×, 200× and 400×. In fact, these magnifications are produced by 4×, 10×, 20×, and 40× objective lens, respectively and a 10× ocular lens. No normalization nor colour standardization was undertaken on the raw images. Each image is in 700×460 pixels dimension with a 3-byte colour depth (3-channels: Red, Green, Blue) and saved in Portable Network Graphics (PNG) format. The BreakHis dataset determines the class of tumours (Benign or Malignant) as well as their type. There are four types of each tumour class. Benign tumour types: Adenosis, Fibroadenoma, Tubular Adenoma and Phyllodes Tumour; Malignant tumour types: Ductal Carcinoma, Lobular Carcinoma, Mucinous Carcinoma (Colloid) and Papillary Carcinoma \cite{spanhol2015dataset}. Some samples of these tumour types are shown in \autoref{table2} with the magnification specific class distribution in \autoref{table1}. For this research we used 70\% of the images in the dataset as training set and the remaining 30\% as test set. \\ \par

\begin{table}[pos=h]
	\caption{Distribution of BreakHis samples.}
	\label{table1}
	\centering
	\begin{tabular*}{\tblwidth}{@{} LLLL@{} }
		\toprule
		\multicolumn{1}{c}{\begin{tabular}[c]{@{}c@{}}BreakHis \\Magnification factor\end{tabular}} & Benign & Malignant & Total  \\ 
		
		\midrule
		\multicolumn{1}{c}{ 40×}      & 652                      & 1370                        & 1995            \\
		\multicolumn{1}{c}{100×}      & 644                      & 1437                        & 2081            \\
		\multicolumn{1}{c}{200×}      & 623                      & 1390                        & 2013            \\
		\multicolumn{1}{c}{400×}      & 588                      & 5429                        & 1820            \\
		\multicolumn{1}{c}{Total}              & 2480                     & 5429                        & 7909            \\
		\bottomrule
	\end{tabular*}
\end{table}

Since BreakHis uses the Histopathology (HP) method which provides coloured images and the possibility of in-depth tissue analysis, there is more chance for early-stage diagnosis of cancer and higher confidence on the results in comparison with other methods. this method is known as an invasive method and requires high care in the process of biopsy \cite{shah2022artificial}. \\

\subsection{Methodology}
Transfer learning is the main idea for the proposed model's binary classification on the BreakHis dataset. Stevo Bozinovski and Ante Fulgosi were the first to mention using transfer learning in a neural network \cite{bozinovski2020reminder}. Their paper was published in 1976, five years later they reported the transfer learning application in training a neural network on a dataset of computer terminal letter images. \par
Deep feature transfer learning is the idea of using a neural network that is trained on a massive dataset such as ImageNet \cite{deng2009imagenet} with hundreds of classes to predict labels on another dataset. This idea could be achieved by removing fully connected layers, which are used to classify images according to features extracted by convolutional layers and adding another classifier to the network, therefore allowing us to use features extracted by a network with a massive dataset with another classification method in different models. The classifier we add to the model can either be a fully connected layer or one based on other classification algorithms. 

\subsubsection{Feature Extractor}
After testing 16 pre-trained neural networks, Inception-ResNet-v2 which is trained on ImageNet showed the most promising results in our case; Accuracies of extracting features by different neural networks are available in \autoref{section4}. The Inception-ResNet-v2 was first introduced by Szegedy et al. in 2016 \cite{szegedy2017inception}. The idea of this neural network consists of residual connections and architecture of Inception. To achieve the advantages of residual connection and its efficiency in computation at the same time, the filter connection stage of Inception architecture was replaced with residual connections. In \cite{szegedy2017inception}, the authors argued that using residual connections causes great improvement in training speed. The architecture of Inception-ResNet-v2 is illustrated in \autoref{fig2}. Every convolution marked by “V” is a valid-padded layer (filter window stays inside the input) and those that are not marked by “V” are same-padded (output window is the same size as the input). There are three inception blocks in the architecture of Inception-ResNet-v2 and in each block, just before matching the depth of input, a 1×1 convolution layer is placed to increase the filter bank dimensionality \cite{szegedy2017inception}.

\subsubsection{Gradient Class Activation Map (Grad-CAM)}
Spatial information is naturally preserved in convolutional layers in general and as a result, it is anticipated that the last convolutional layer of the feature extractor has the best balance between high-level semantics and detailed spatial information. Neurons in the last convolutional layer search for semantic class-specific data in the given image \cite{selvaraju2017grad}. Gradient information streaming into these layers is used by Grad-CAM to appoint significance values to each neuron for a specific decision of interest, thus emphasizing the identification area for each class in the input image \cite{selvaraju2017grad, kim2021acoustic}. In other words, Grad-CAM determines the gradient of a particular class grade with regards to the pixel of the feature map and then averages gradients of each channel to come by channel-wise weight \cite{si2021spatial}. With this information, it is possible for us to disclose the internal logic of CNNs as well as provide clinical references about patients' data and health status to a greater extent \cite{xie2021interpretable, nasiri2021novel, nasiri2022automated}. We can also regard these visual insights to determine where the model is failing and why then address these problems by changing the model's architecture \cite{neal20213d}. Examples of applying grad-cam on features extracted by Inception-ResNet-v2 are shown in \autoref{table3}.

\begin{table*}[width=1\textwidth,cols=4,pos=!h]
	\caption{Inception-ResNet-v2 Grad-CAM visualization samples.}
	\label{table3}
	\begin{tabular*}{\tblwidth}{@{} LLLL@{} }
		\toprule
		\multicolumn{1}{c}{Tumour class} & \multicolumn{1}{c}{Original biopsy image} & \multicolumn{1}{c}{Heatmap} & \multicolumn{1}{c}{Grad-CAM} \\ 
		
		\midrule
		\multicolumn{1}{c|}{\multirow{9}{*}{Benign}}    & \begin{minipage}{.22\textwidth} \centering\includegraphics[height=25mm]{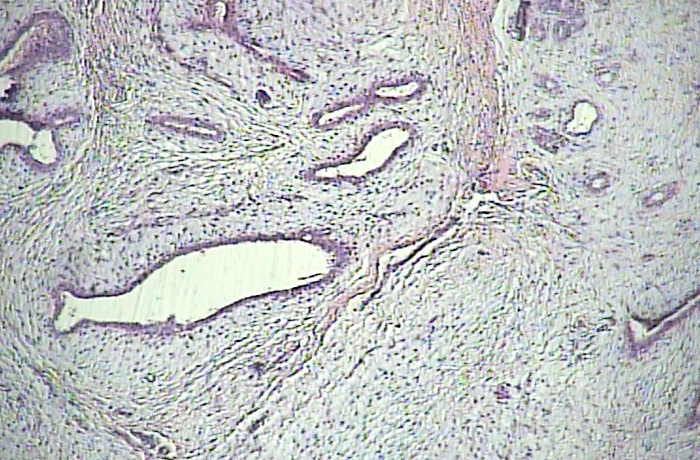} \end{minipage} & \begin{minipage}{.22\textwidth} \centering\includegraphics[height=25mm]{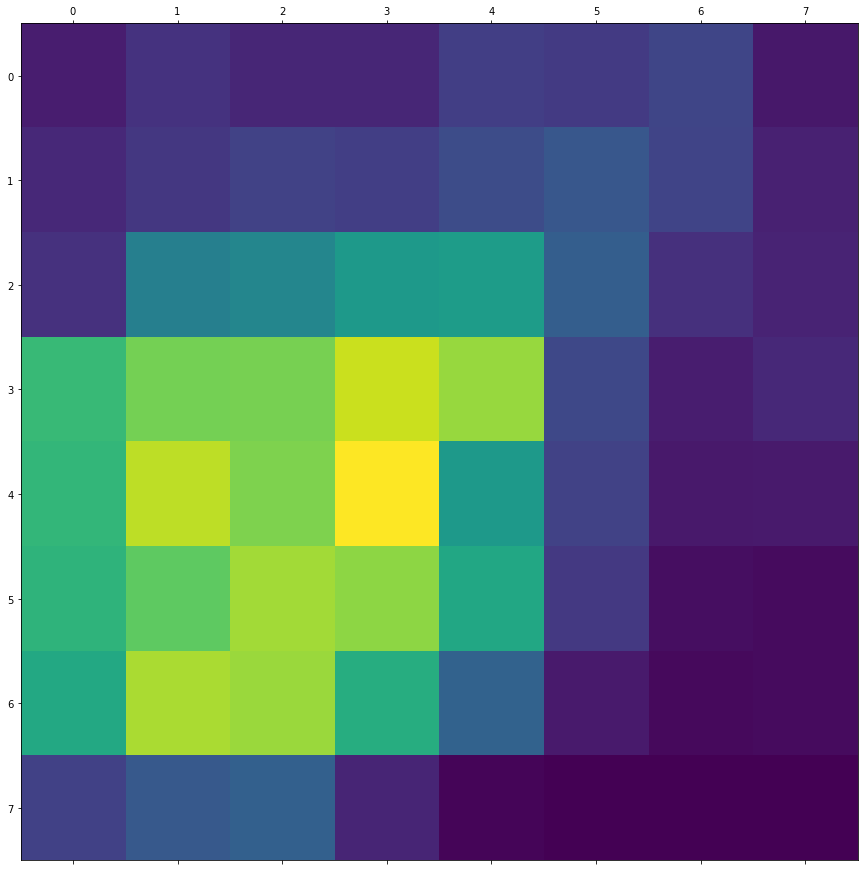} \end{minipage} & \begin{minipage}{.22\textwidth} \centering\includegraphics[height=25mm]{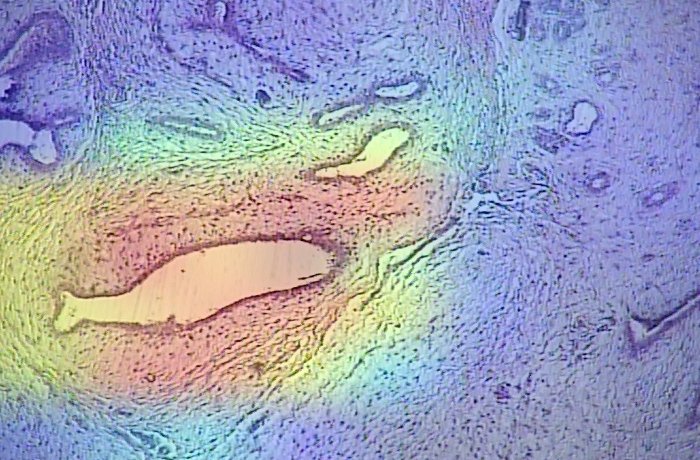} \end{minipage} \\
		\multicolumn{1}{c|}{}                           & & & \\
		\multicolumn{1}{c|}{}                           & \begin{minipage}{.22\textwidth} \centering\includegraphics[height=25mm]{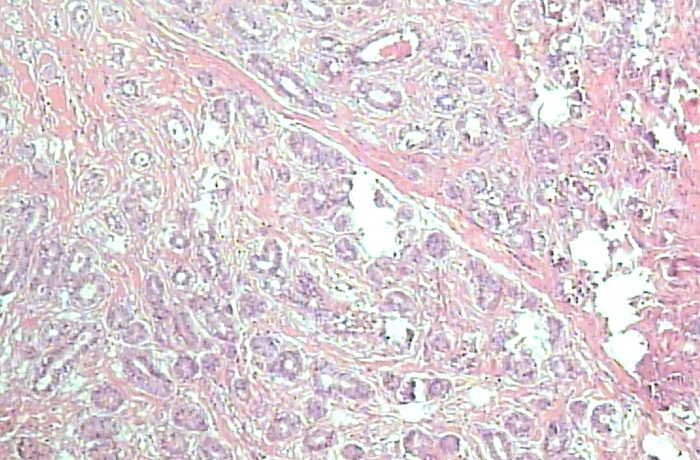} \end{minipage} & \begin{minipage}{.22\textwidth} \centering\includegraphics[height=25mm]{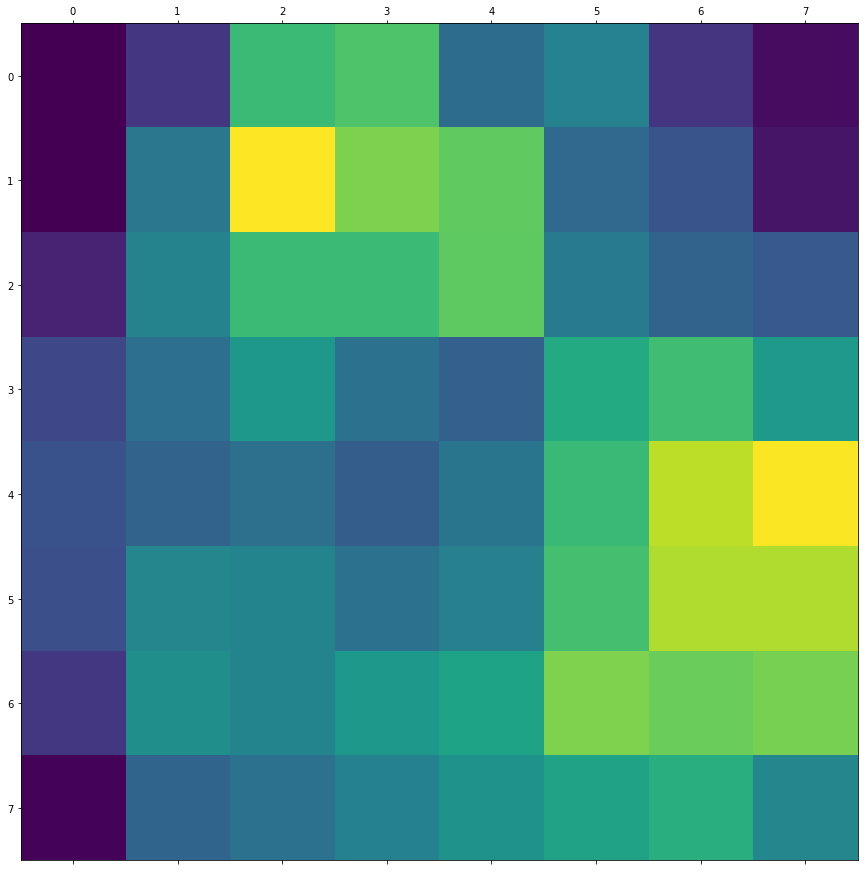} \end{minipage} & \begin{minipage}{.22\textwidth} \centering\includegraphics[height=25mm]{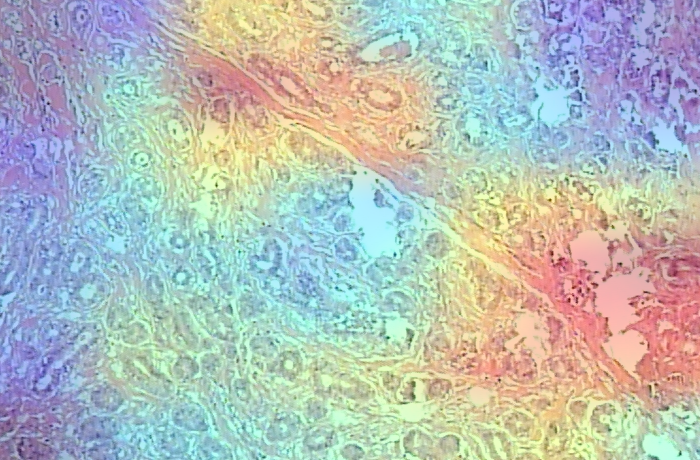} \end{minipage} \\
		
		\multicolumn{1}{c}{}                            & & & \\
		
		\multicolumn{1}{c|}{\multirow{9}{*}{Malignant}}	& \begin{minipage}{.22\textwidth} \centering\includegraphics[height=25mm]{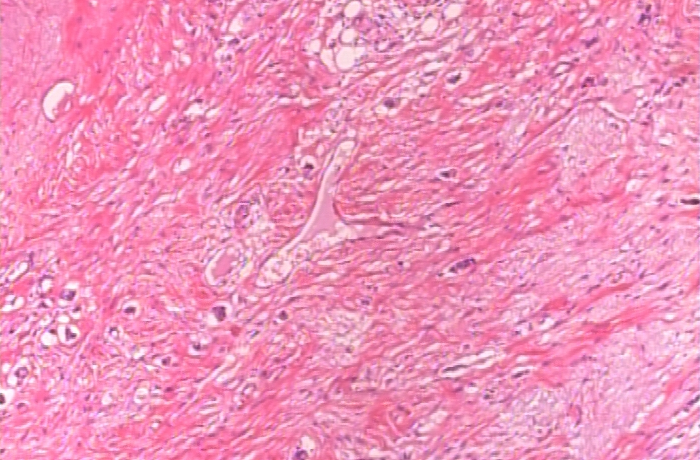} \end{minipage} & \begin{minipage}{.22\textwidth} \centering\includegraphics[height=25mm]{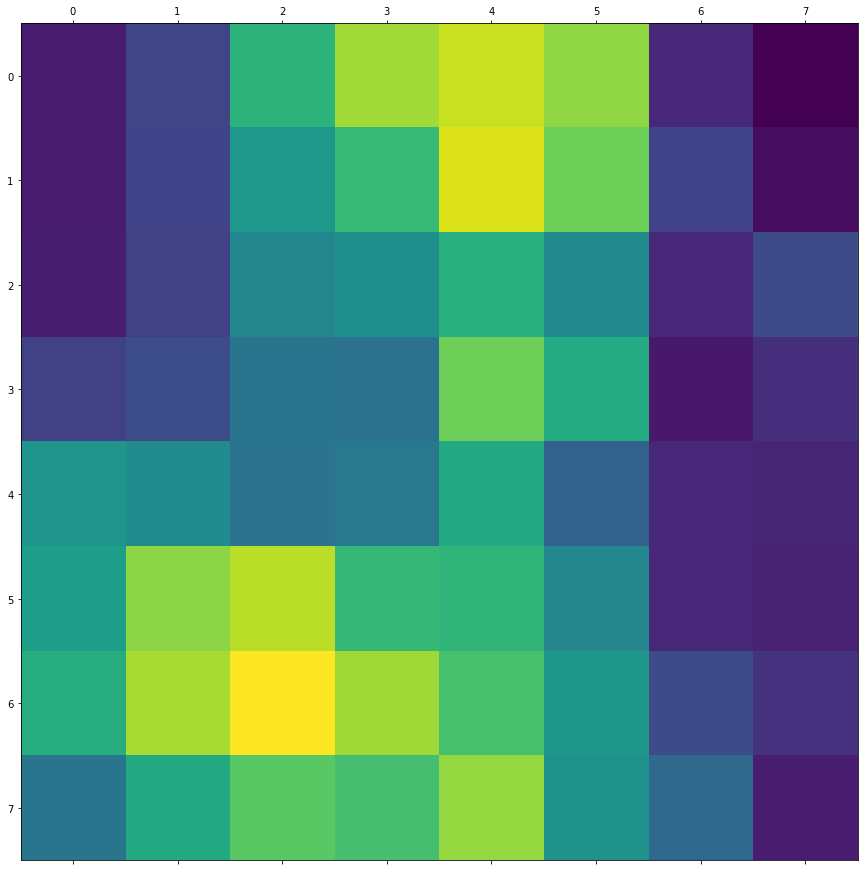} \end{minipage} & \begin{minipage}{.22\textwidth} \centering\includegraphics[height=25mm]{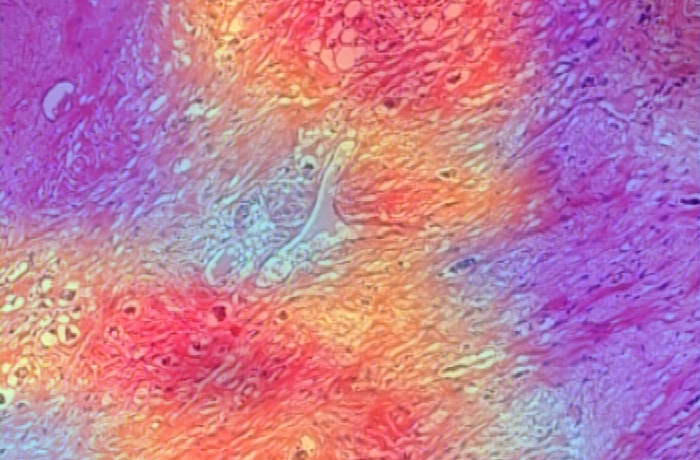} \end{minipage} \\
		\multicolumn{1}{c|}{}                           & & & \\
		\multicolumn{1}{c|}{}                           & \begin{minipage}{.22\textwidth} \centering\includegraphics[height=25mm]{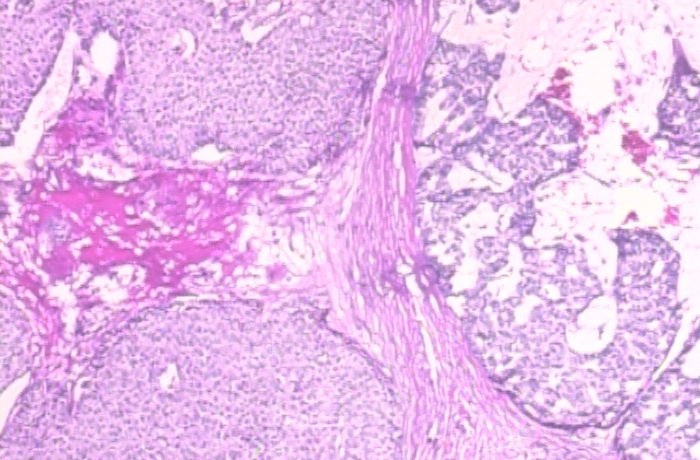} \end{minipage} & \begin{minipage}{.22\textwidth} \centering\includegraphics[height=25mm]{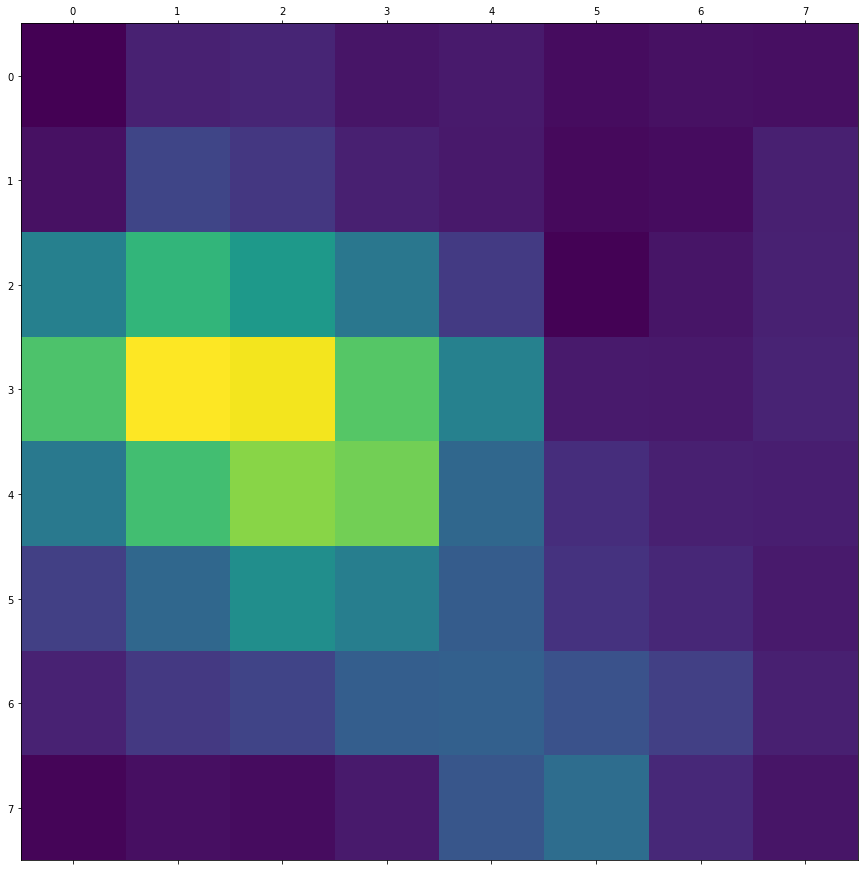} \end{minipage} & \begin{minipage}{.22\textwidth} \centering\includegraphics[height=25mm]{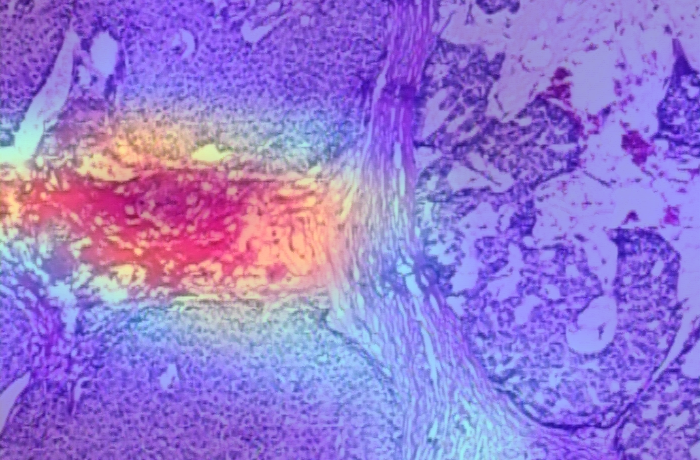} \end{minipage} \\
		\bottomrule
	\end{tabular*}
\end{table*}

\subsubsection{Classifier}
Three different classifiers were used in a variety of different model structures in this study to determine whether a tumour is benign or malignant based on the deep features extracted by the deep neural network, however, we will only discuss the model that lent to the best results and only mention the others. XGBoost, CatBoost and LightGBM were used alone in three of the seven models while the other four used a soft voting ensemble method of two or all three of the mentioned classifiers instead. Best results were achieved by LightGBM, XGBoost and CatBoost classifier’s soft voting which is described in \ref{section324}.

\begin{figure}[!h]
	\centering           
	\includegraphics[clip, trim = 10pt  5cm  1cm  2cm, width= .46\textwidth]{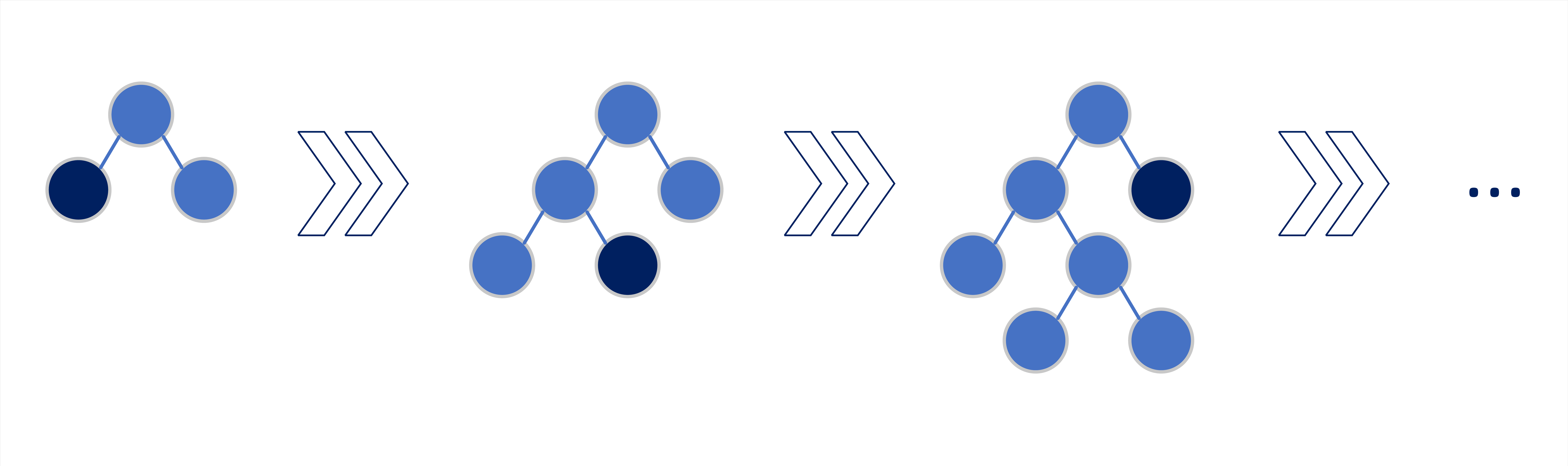}
	\caption{LightGBM leaf-wise tree growth.}
	\label{fig3}
\end{figure}

\paragraph{\textbf{LightGBM:}} 
Gradient boosting decision tree (GBDT) is a time-honoured model that utilizes a weak classifier, namely a decision tree, with iterative training to achieve a top-notch model that enjoys such advantages as competent training effect and resistance to overfitting \cite{xu2021intelligent}.LightGBM (light gradient boosting machine) is a speedy, open-source lifting framework based on GBDT which has a faster training speed, lower memory consumption, and better accuracy \cite{sawant2021automated,nasiri2022classification}. LightGBM can quickly process massive data by instigating leaf-wise tree growth (see \autoref{fig3}) with depth limitation and gradient-based one-side sampling as well as exclusive feature bundling approaches \cite{xu2021intelligent, ezzoddin2022diagnosis}. LightGBM hyperparameters were tuned by trial and error. 

\paragraph{\textbf{CatBoost:}} 
This algorithm integrates GBDT and categorical features hence the name CatBoost. CatBoost is oblivious tree-based with fewer parameters and categorical variable support. The mentioned implementation also handles gradient bias and prediction shift, doing so improves the generalization and robustness of the algorithm \cite{zhou2021fire}. CatBoost (categorical boosting) uses permutation, target-based statistics and one-hot-max-size (OHMS) to emphasize categorical columns while utilizing the greedy algorithm at each new tree split to resolve the exponential feature combination increase \cite{al2019comparison}. CatBoost boasts the characteristics of high classification accuracy and fast training speed \cite{wang2020fuzzy}. In fact, the highest single magnification classification accuracy result (i.e., 97.49\%) was achieved by using CatBoost as the model’s classifier however the ensemble method worked better across all magnifications in average. In this study, CatBoost hyperparameters were tuned by trial and error.

\begin{figure}[!h]
	\centering           
	\includegraphics[clip, trim = 10pt  6cm  1cm  2cm, width=.46\textwidth]{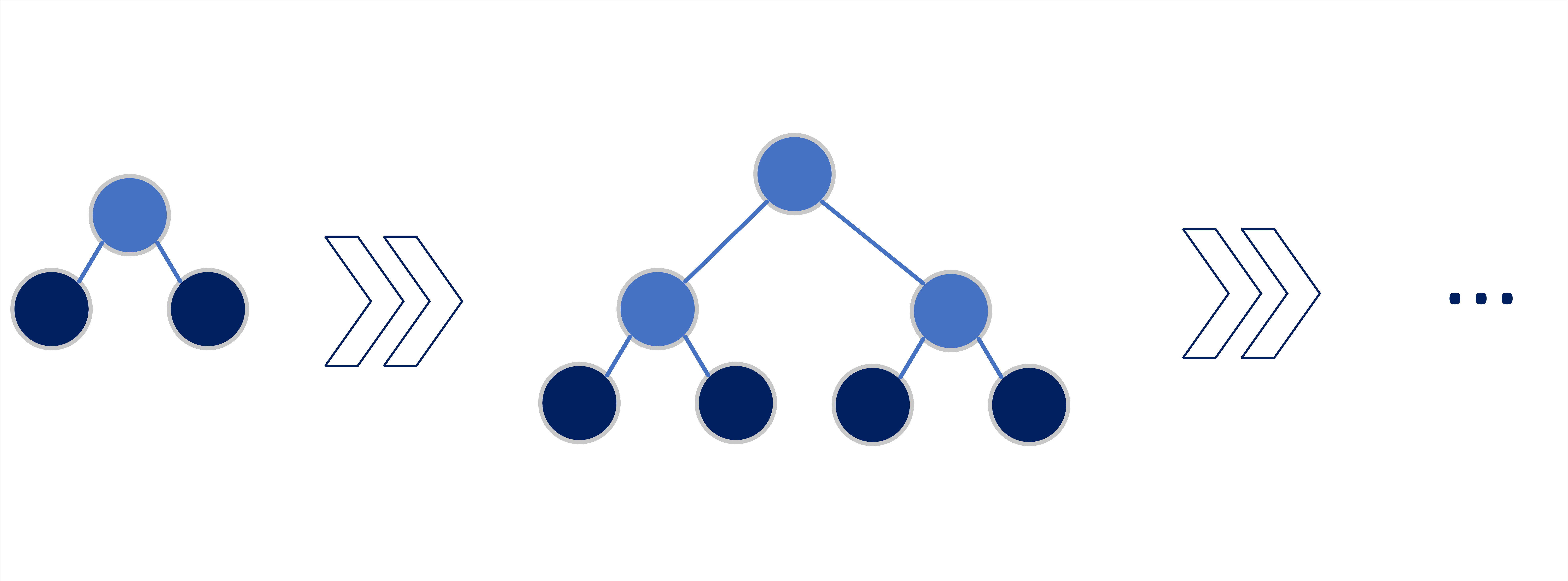}
	\caption{XGBoost level-wise tree growth.}
	\label{fig4}
\end{figure}

\begin{figure*}[!p]
	\centering           
	\includegraphics[clip, trim = 5cm  20cm  5cm  4cm, width=\textwidth]{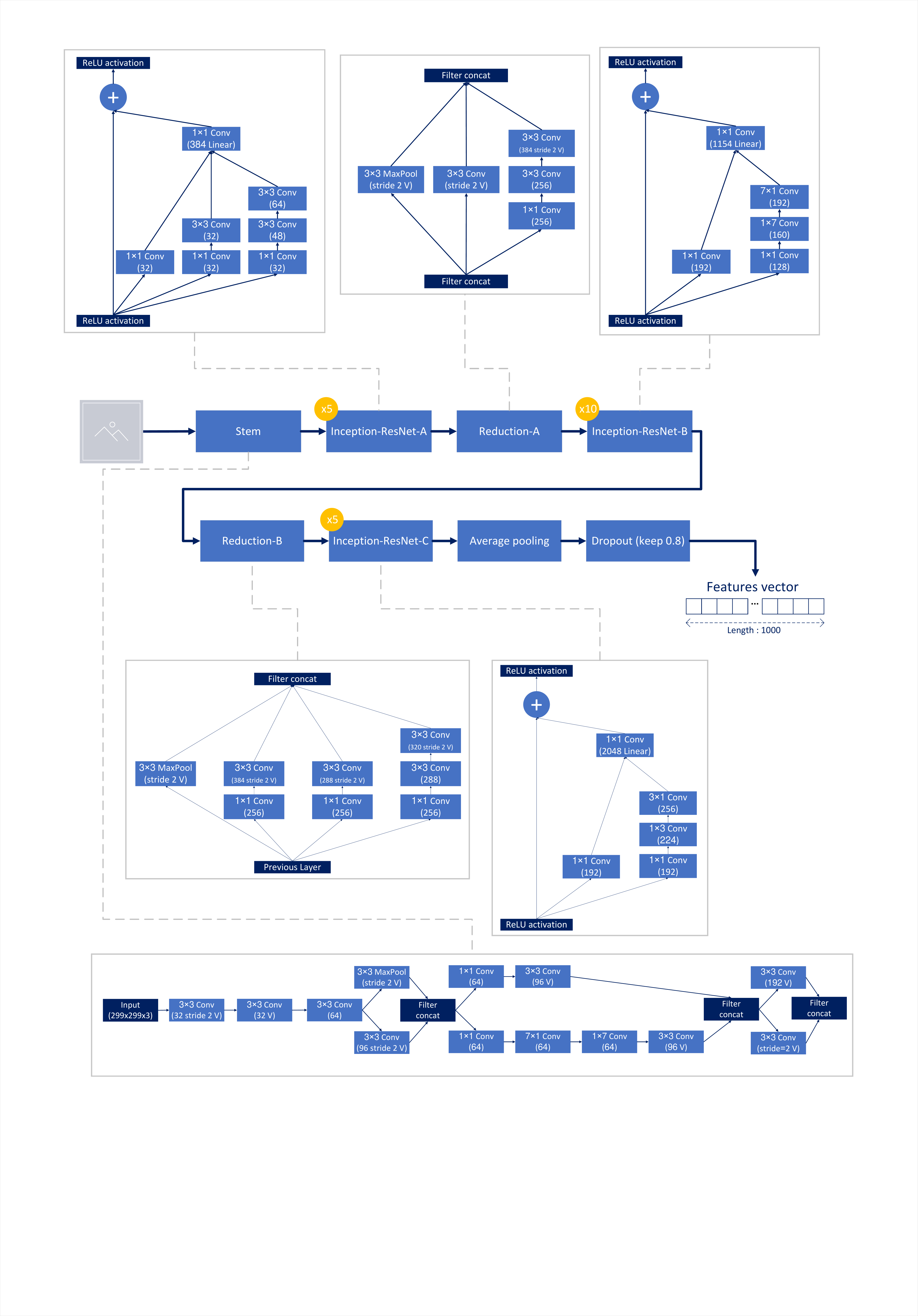}
	\caption{Architecture of Inception-ResNet-v2 as feature extractor.}
	\label{fig2}
\end{figure*}

\paragraph{\textbf{XGBoost:}} 
Extreme Gradient Boosting (XGBoost) \cite{chen2016xgboost}, like the other two classifiers used in the proposed majority voting ensemble method, is a technique that uses the results of several decision trees to generate a prediction score \cite{chelgani2021interpretable}. This method expands the tree level-wise (see \autoref{fig4}) by constantly dividing features. Every tree then computes the feature and threshold with the best branch effect and proceeds to complete the split. Simply put, XGBoost yields better results than conventional decision trees by utilizing classification and regression trees as base learners to consecutively blend various tree predictions with the use of gradient boosting as an error minimizer \cite{nasiri2021novel}. XGBoost also lowers the likelihood of over-fitting through regularization by employing a second–order Taylor series to approximate the value of the loss function \cite{chelgani2021interpretable, chelgani2021modeling, akhavan2021internet}. All of this makes XGBoost a gradient boosting library optimized for efficiency, versatility and portability \cite{nasiri2021prediction,hasani2022cov}. XGBoost hyperparameters were tuned by trial and error.

\subsubsection{Ensemble Classification}
\label{section324}
In this study, three fine-tuned classifiers are used to predict the probability of malignancy based on features vector obtained from pre-trained CNNs. 
We decided to implement different combinations of these classifiers, which use a form of ensemble learning themselves, by utilizing another ensemble method. 
There are multiple ways to adopt ensemble learning in classification, but here we use one of the most common ones called Soft-Voting. In this approach, every classifier produces a probability for each class, the mean of which is considered as the final probability of that class \cite{gupta2020improving}. Finally, the class with the highest probability will be the agreement of ensembled classifiers. This approach has no limitation on the number of classifiers while another method like Majority-Voting can only ensemble an even number of classifiers. 
\\ ~ \\ ~

\section{Result and Discussion}
\label{section4}
\subsection{Evaluation Metrics}
In our binary classification of the BreakHis dataset, predicted labels can be split into four different categories by comparing them to actual labels of test set: TP (number of correct malignant predictions), TN (number of correct benign predictions), FP (number of incorrect malignant predictions), FN (number of incorrect benign predictions). The following are different evaluation metrics used in this paper:\begin{flalign}
	& Accuracy =  \frac{TP+TN}{TP+FP+TN+FN} &\\ \nonumber \\
	& Balanced~accuracy =  \frac{1}{2}(\frac{TP}{TP+FN} + \frac{TN}{TN+FP}) &\\ \nonumber \\
	& F_1 score = \frac{TP}{TP + \frac{1}{2} (FP+FN)} &
\end{flalign}

\newpage
\subsection{Binary Classification Results}
In this paper, 16 pre-trained networks were used to extract features from input images. Features were then supplied to seven different classifiers (three single and four soft voting ensembles) to predict our final labels. Results of the performance of individual and ensemble models on the test set are available in \autoref{table4}; X, L and C stand for XGBoost, LightGBM and CatBoost, respectively. The Keras implementation of the pre-trained networks (CNN) is used in this research. For each network, a simple preprocessing method recommended by Keras is used before feeding images to the CNNs \footnote{\url{https://github.com/keras-team/keras/tree/master/keras/applications}, Retrieved on 2021-8-5}.
The Google-Colaboratory free environment (12GB RAM, 78GB disk and cloud GPU) has been used for models implementation.

\subsection{Proposed Method}
\begin{figure}[!h]
	\centering           
	\includegraphics[clip, trim = 4cm  1cm  4cm  1cm, width=.5\textwidth]{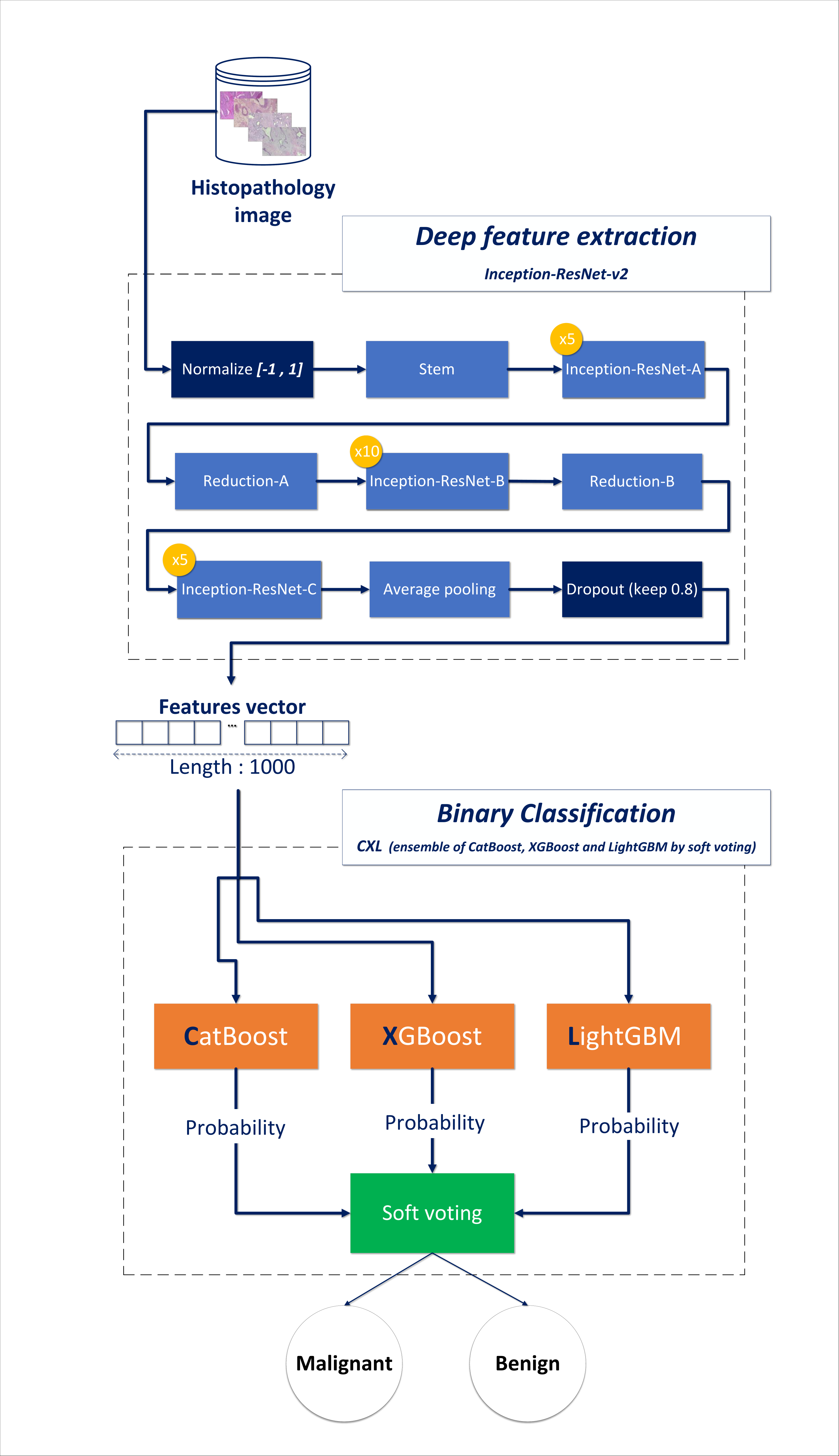}
	\caption{Proposed method (IRv2-CXL) architecture.}
	\label{fig5}
\end{figure}

\begin{figure*}[!h]
	\centering
	\begin{subfigure}[t]{0.45\textwidth}
		\includegraphics[height=75mm]{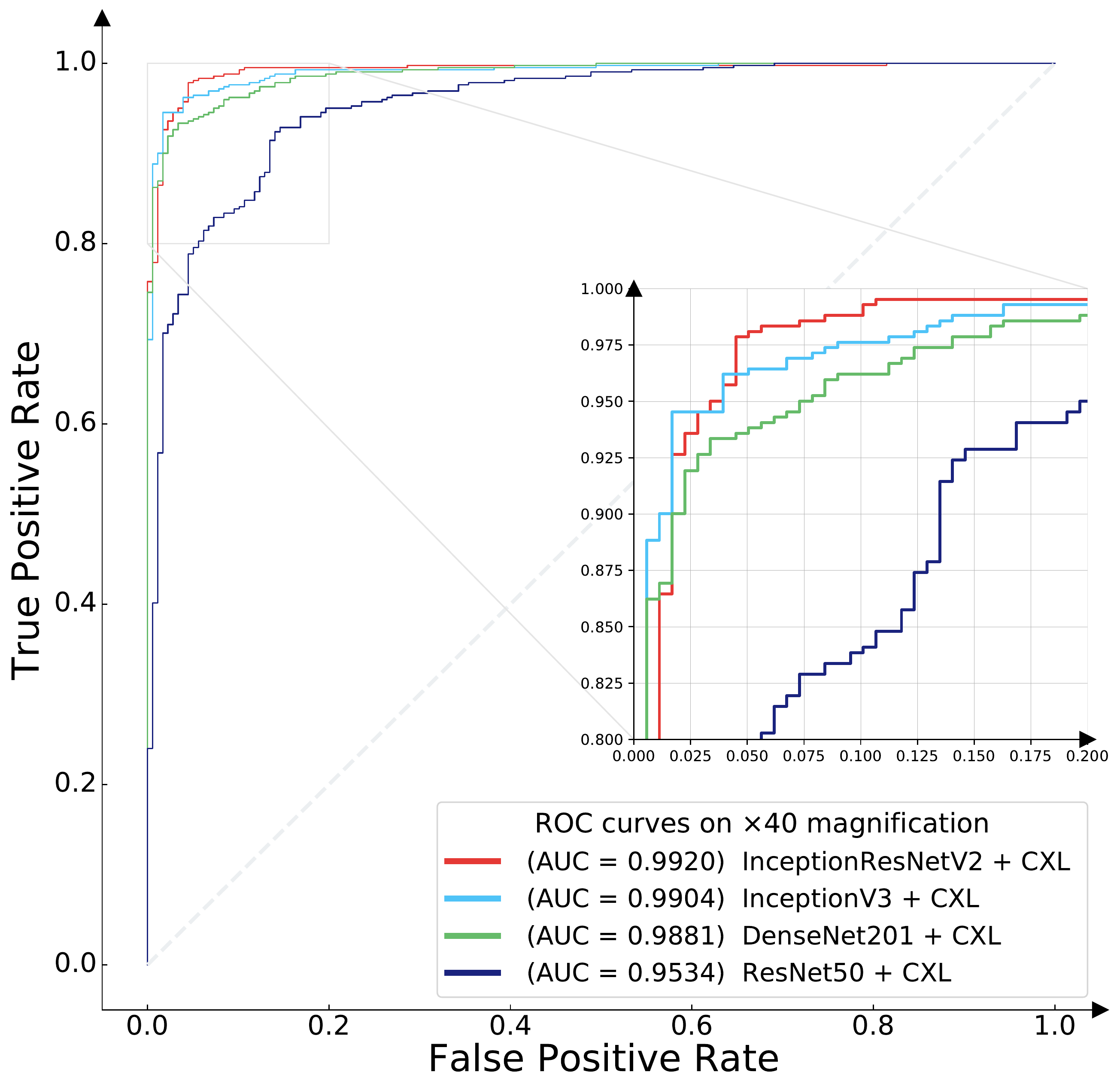}
		\caption{}
	\end{subfigure}
	\hfill
	\begin{subfigure}[t]{0.45\textwidth}
		\includegraphics[height=75mm]{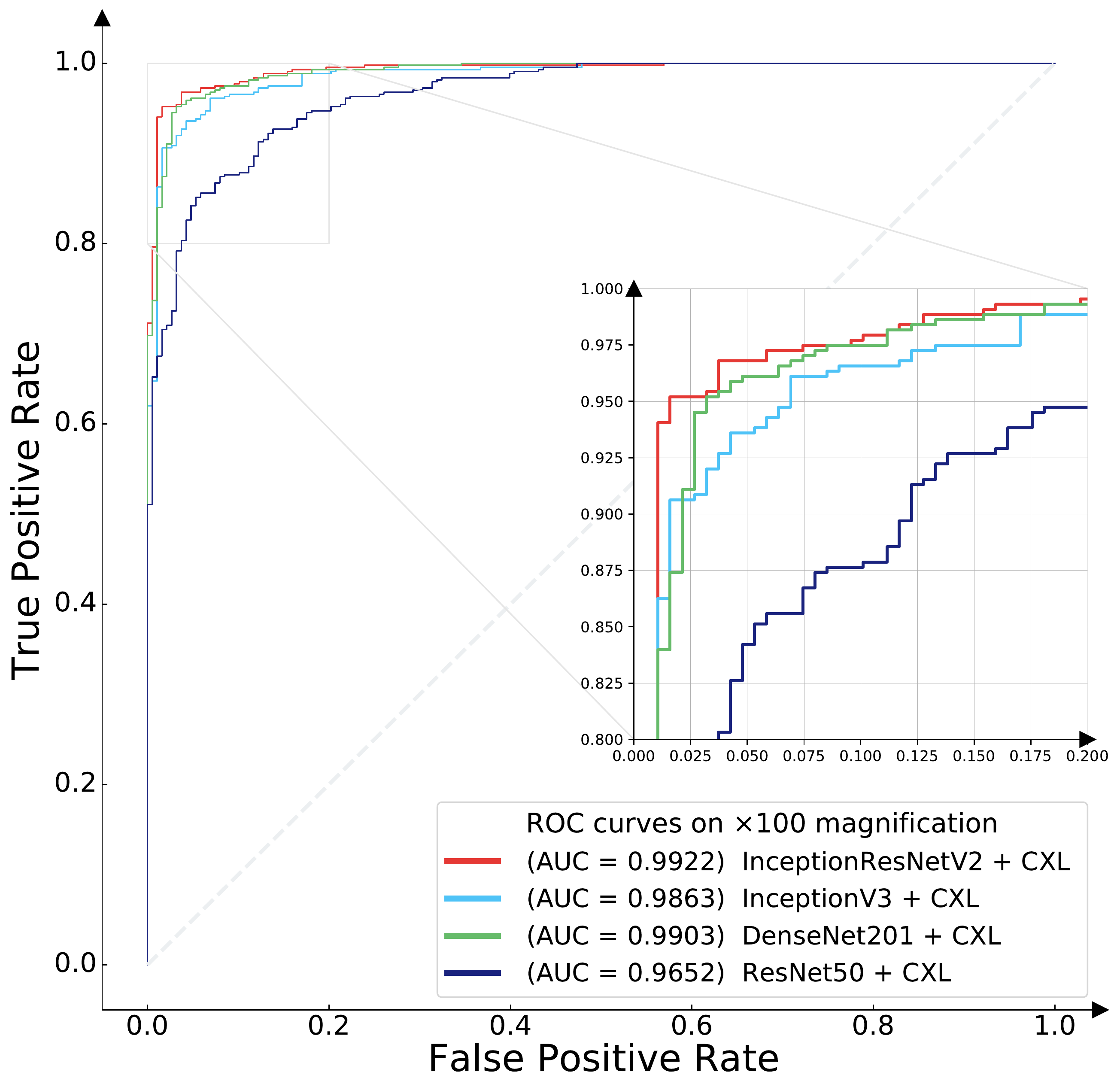}
		\caption{}
	\end{subfigure}
	\par\bigskip
	\begin{subfigure}[t]{0.45\textwidth}
		\includegraphics[height=75mm]{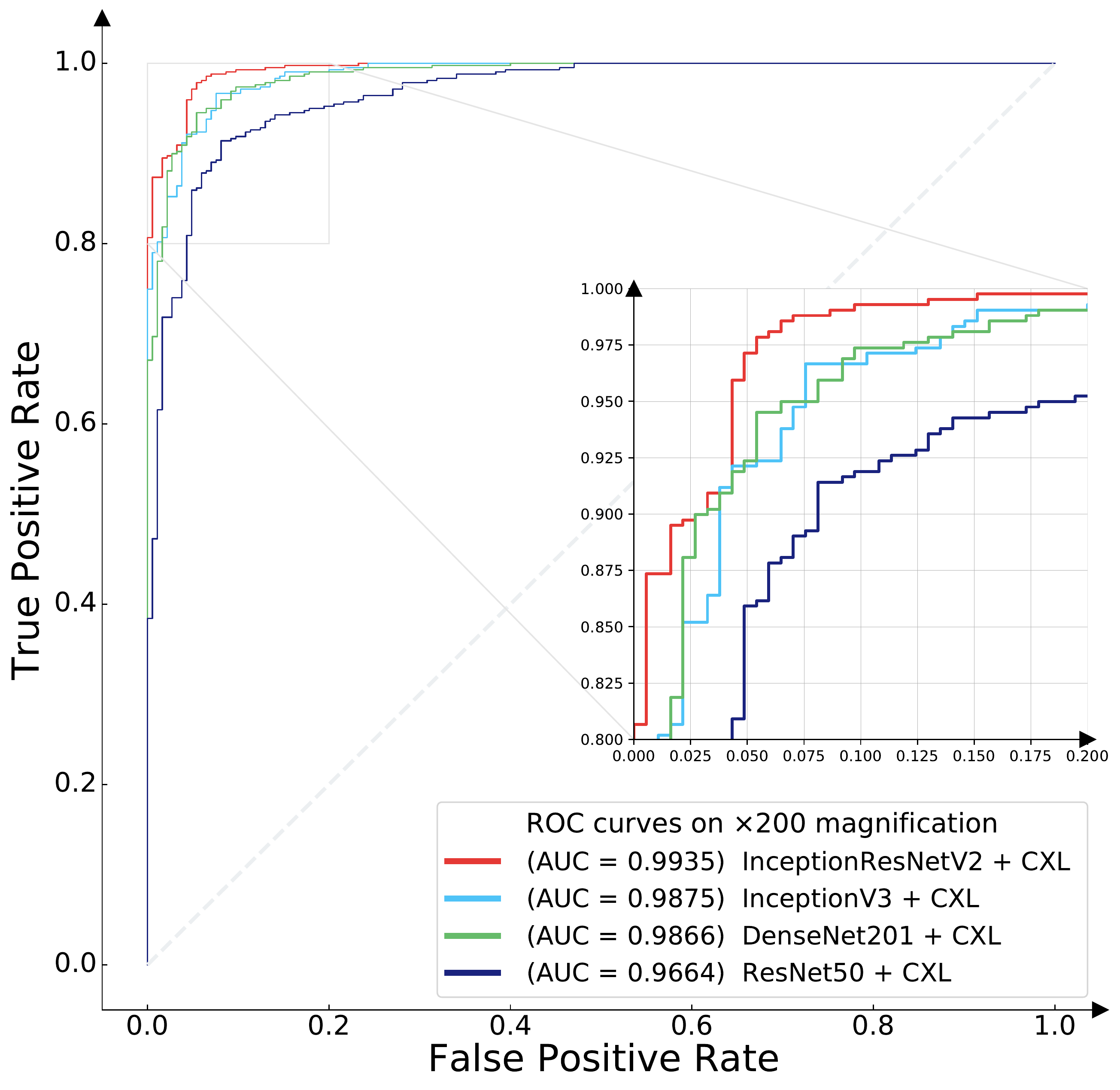}
		\caption{}
	\end{subfigure}
	\hfill
	\begin{subfigure}[t]{0.45\textwidth}
		\includegraphics[height=75mm]{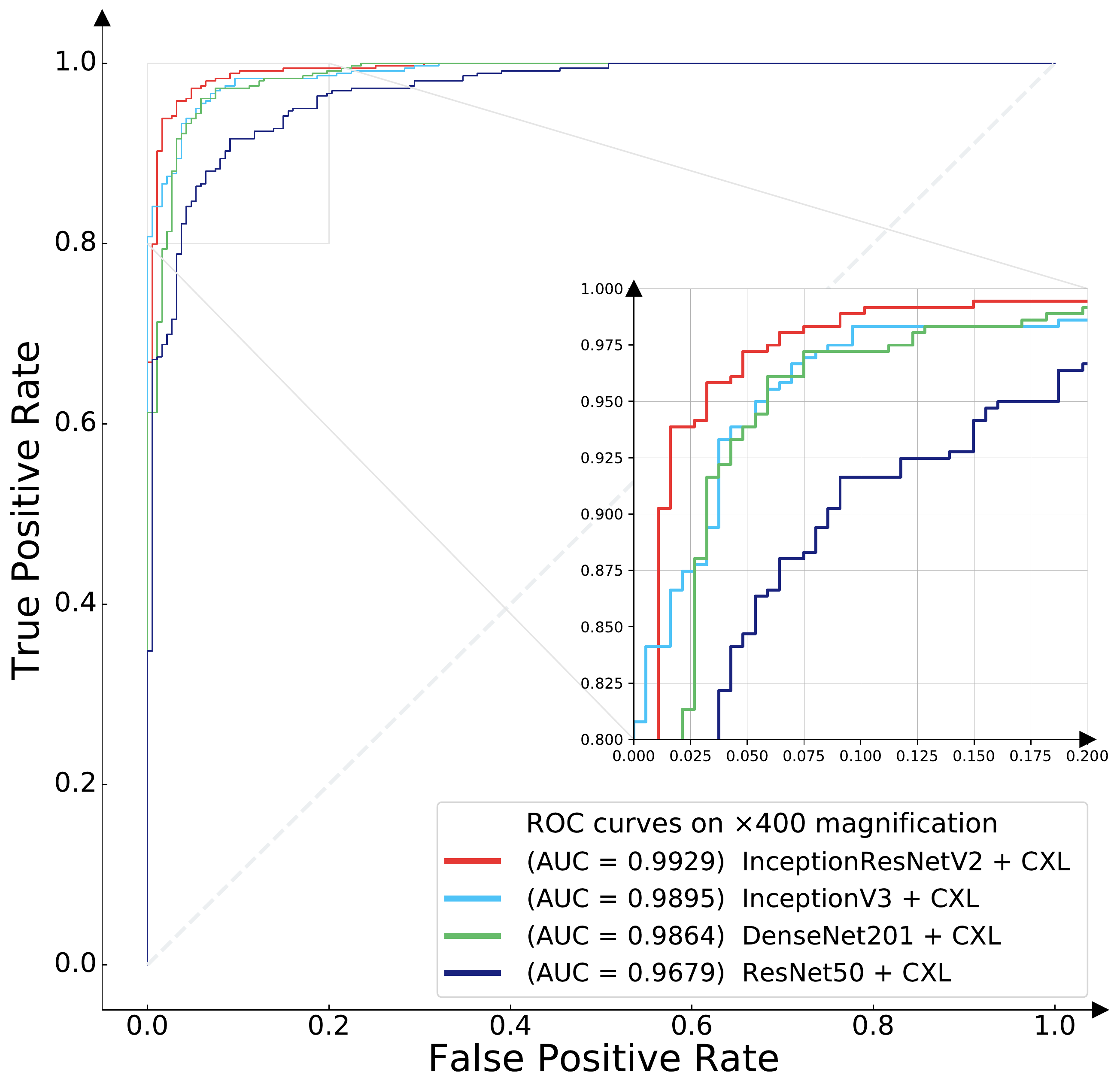}
		\caption{}
	\end{subfigure}
	\caption{ROC curves of different feature extractors classified by CXL. (a) 40×, (b) 100×, (c) 200×, (d) 400×.}
	\label{table5}
\end{figure*}

\begin{figure*}[!h]
	\centering
	\begin{subfigure}[t]{0.2\textwidth}
		\includegraphics {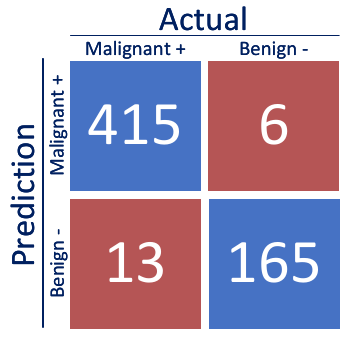}
		\caption{}
	\end{subfigure}
	\qquad
	\begin{subfigure}[t]{0.2\textwidth}
		\includegraphics {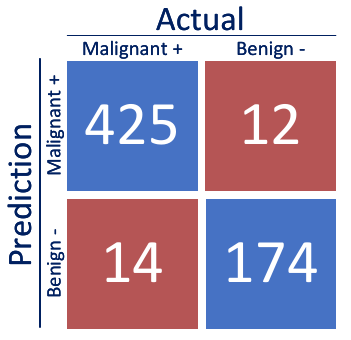}
		\caption{}
	\end{subfigure}
	\qquad
	\begin{subfigure}[t]{0.2\textwidth}
		\centering
		\includegraphics {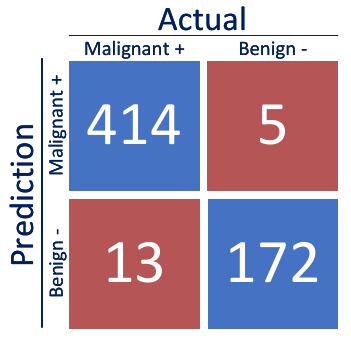}
		\caption{}
	\end{subfigure}
	\qquad
	\begin{subfigure}[t]{0.2\textwidth}
		\centering
		\includegraphics {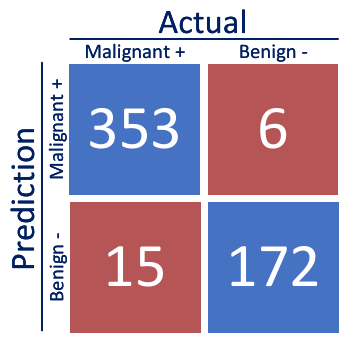}
		\caption{}
	\end{subfigure}
	\caption{IRv2-CXL confusion matrix on different test set magnifications.  (a) 40×, (b) 100×, (c) 200×, (d) 400×.}
	\label{table7}
\end{figure*}

As evident by the results presented in \autoref{table4}, the combination of Inception-ResNet-v2 as feature extractor and an ensemble of all three classifiers produced the best accuracy among all 112 tested models (results with the f1-score as an evaluation metric are also available in \autoref{table9}), so it was selected as the proposed method of this study. It consists of Inception-ResNet-v2 and ensemble of CatBoost, XGBoost, and LightGBM, which we named IRv2-CXL. The full architecture of IRv2-CXL is illustrated in \autoref{fig5}. By reviewing Benhammou \cite{benhammou2020breakhis} and the other studies mentioned in related works, we concluded that ResNet50, DensNet201 and Inception-v3 networks are three of the more commonly used pre-trained networks in breast tumours classification. Therefore, we decided to compare the ROC curve of these networks with Inception-ResNet-v2 using the chosen classifier (i.e., CXL). Results are shown in \autoref{table5}. Other evaluation metrics applied to the proposed method are provided in \autoref{table6} and the magnification specific confusion matrix in \autoref{table7}.

\begin{table}[pos = h]
	\caption{Proposed method (IRv2-CXL) evaluation.}
	\label{table6}
	\begin{tabular}{@{}cccccc@{}}
		\toprule
		\multirow{2}{*}{Metric}                                        & \multicolumn{5}{c}{Results}                              \\ \cmidrule(l){2-6} 
		& 40×     & 100×    & 200×    & 400×    & \textbf{Average} \\
		\midrule                                                              
		$Accuracy$                                                       & 96.83\% & 95.84\% & 97.01\% & 96.15\% & \textbf{96.46\%} \\
		&&&&\\
		\begin{tabular}[c]{@{}c@{}}$Balanced$  \\  $accuracy$\end{tabular} & 95.64\% & 94.90\% & 95.89\% & 95.15\% & \textbf{95.40\%} \\
		&&&&\\
		$F_1 score$                                                       & 97.76\% & 97.03\% & 97.87\% & 97.11\% & \textbf{97.44\%} \\
		\bottomrule
	\end{tabular}%
\end{table}

\begin{table*}[width=.99\textwidth,cols=4,pos=!h]
	\caption{Binary classification \textbf{accuracy} results on BreakHis dataset.}
	\label{table4}
	\centering
	\footnotesize
	\begin{tabular*}{\textwidth}{@{}llccccp{1cm}llcccc@{}}
		\toprule
		\multicolumn{1}{c}{\multirow{2}{*}{\begin{tabular}[c]{@{}c@{}}Feature \\ extractor\end{tabular}}} & \multicolumn{1}{c}{\multirow{2}{*}{Classifier(s)}} & \multicolumn{4}{c}{Result   (accuracy \%)} &  & \multicolumn{1}{c}{\multirow{2}{*}{\begin{tabular}[c]{@{}c@{}}Feature \\ extractor\end{tabular}}} & \multicolumn{1}{c}{\multirow{2}{*}{Classifier(s)}} & \multicolumn{4}{c}{Result   (accuracy \%)} \\ \cmidrule(lr){3-6} \cmidrule(l){10-13} 
		\multicolumn{1}{c}{}                                                                             & \multicolumn{1}{c}{}                               & 40×       & 100×     & 200×     & 400×     &  & \multicolumn{1}{c}{}                                                                             & \multicolumn{1}{c}{}                               & 40×       & 100×     & 200×     & 400×     \\ \cmidrule(r){1-6} \cmidrule(l){8-13} 
		\multirow{7}{*}{VGG 16}                                                                          & X                                                  & 85.64     & 81.44    & 83.44    & 78.20    &  & \multirow{7}{*}{\begin{tabular}[c]{@{}l@{}}Inception\\ v3\end{tabular}}                          & X                                                  & 94.49     & 93.6     & 94.7     & 93.58    \\
		& L                                                  & 85.97     & 82.56    & 85.26    & 78.38    &  &                                                                                                  & L                                                  & 95.82     & 94.56    & 94.53    & 94.68    \\
		& C                                                  & 86.47     & 83.68    & 85.59    & 78.93    &  &                                                                                                  & C                                                  & 95.15     & 93.44    & 93.70    & 94.13    \\
		& X \& L                                             & 87.31     & 82.88    & 84.76    & 79.48    &  &                                                                                                  & X \& L                                             & 94.99     & 93.92    & 94.86    & 95.23    \\
		& X \& C                                             & 86.64     & 82.24    & 84.76    & 79.30    &  &                                                                                                  & X \& C                                             & 95.15     & 93.76    & 94.70    & 94.87    \\
		& L \& C                                             & 86.31     & 84.00    & 85.09    & 79.48    &  &                                                                                                  & L \& C                                             & 95.32     & 94.08    & 94.70    & 94.87    \\
		& X \& L \& C                                        & 86.97     & 83.04    & 84.76    & 78.93    &  &                                                                                                  & X \& L \& C                                        & 95.49     & 94.24    & 94.53    & 95.23    \\ \cmidrule(r){1-6} \cmidrule(l){8-13} 
		\multirow{7}{*}{VGG 19}                                                                          & X                                                  & 85.64     & 80.32    & 83.27    & 78.38    &  & \multirow{7}{*}{ \begin{tabular}[c]{@{}l@{}} Inception \\ ResNet \\ v2 \end{tabular}}                 & X                                                  & 96.32     & 95.36    & 95.36    & 95.97    \\
		& L                                                  & 84.47     & 83.04    & 83.6     & 78.20    &  &                                                                                                  & L                                                  & 96.82     & 95.52    & 96.85    & 96.33    \\
		& C                                                  & 85.97     & 82.72    & 85.43    & 79.67    &  &                                                                                                  & C                                                  & \textbf{97.49}     & 95.84    & 96.19    & 95.97    \\
		& X \& L                                             & 85.64     & 81.6     & 85.43    & 80.58    &  &                                                                                                  & X \& L                                             & 96.49     & 95.84    & 96.68    & 96.52    \\
		& X \& C                                             & 86.81     & 81.92    & 85.59    & 80.4     &  &                                                                                                  & X \& C                                             & 96.82     & 96.16    & 96.35    & 95.78    \\
		& L \& C                                             & 85.14     & 82.56    & 84.60    & 79.67    &  &                                                                                                  & L \& C                                             & 96.99     & 96.16    & 96.68    & 95.42    \\
		& X \& L \& C                                        & 85.80     & 82.40    & 85.92    & 80.03    &  &                                                                                                  & X \& L \& C                                        & \textbf{96.82}     & \textbf{95.84}    & \textbf{97.01}    & \textbf{96.15}    \\ \cmidrule(r){1-6} \cmidrule(l){8-13} 
		\multirow{7}{*}{\begin{tabular}[c]{@{}l@{}}ResNet \\ 50\end{tabular}}                            & X                                                  & 87.81     & 89.6     & 90.39    & 90.65    &  & \multirow{7}{*}{\begin{tabular}[c]{@{}l@{}}DenseNet \\ 121\end{tabular}}                         & X                                                  & 93.99     & 95.2     & 93.54    & 95.23    \\
		& L                                                  & 89.81     & 89.92    & 90.72    & 90.10    &  &                                                                                                  & L                                                  & 94.15     & 95.84    & 94.53    & 95.23    \\
		& C                                                  & 89.14     & 90.08    & 91.05    & 89.92    &  &                                                                                                  & C                                                  & 94.15     & 94.72    & 93.87    & 94.50    \\
		& X \& L                                             & 89.81     & 90.56    & 91.05    & 90.29    &  &                                                                                                  & X \& L                                             & 94.65     & 96.16    & 94.37    & 94.87    \\
		& X \& C                                             & 89.31     & 90.56    & 91.39    & 90.29    &  &                                                                                                  & X \& C                                             & 93.99     & 95.20    & 94.03    & 94.87    \\
		& L \& C                                             & 90.15     & 90.08    & 90.89    & 90.84    &  &                                                                                                  & L \& C                                             & 93.99     & 95.36    & 94.37    & 95.42    \\
		& X \& L \& C                                        & 89.48     & 90.72    & 91.22    & 90.84    &  &                                                                                                  & X \& L \& C                                        & 94.65     & 95.20    & 95.03    & 95.23    \\ \cmidrule(r){1-6} \cmidrule(l){8-13} 
		\multirow{7}{*}{\begin{tabular}[c]{@{}l@{}}ResNet\\ 101\end{tabular}}                            & X                                                  & 86.14     & 86.4     & 87.58    & 87.36    &  & \multirow{7}{*}{\begin{tabular}[c]{@{}l@{}}DenseNet \\ 169\end{tabular}}                         & X                                                  & 94.65     & 96.00    & 94.20    & 94.68    \\
		& L                                                  & 86.47     & 88.16    & 89.4     & 88.09    &  &                                                                                                  & L                                                  & 94.65     & 96.64    & 94.20    & 93.77    \\
		& C                                                  & 87.14     & 88.00    & 90.06    & 89.56    &  &                                                                                                  & C                                                  & 93.15     & 96.16    & 94.86    & 94.50    \\
		& X \& L                                             & 86.14     & 87.68    & 88.90    & 87.72    &  &                                                                                                  & X \& L                                             & 94.65     & 95.84    & 94.20    & 94.13    \\
		& X \& C                                             & 86.97     & 88.00    & 89.56    & 88.46    &  &                                                                                                  & X \& C                                             & 93.99     & 95.68    & 94.20    & 94.68    \\
		& L \& C                                             & 87.31     & 88.64    & 89.73    & 88.82    &  &                                                                                                  & L \& C                                             & 94.15     & 96.48    & 94.70    & 93.95    \\
		& X \& L \& C                                        & 86.64     & 88.32    & 89.56    & 88.46    &  &                                                                                                  & X \& L \& C                                        & 94.32     & 96.00    & 94.70    & 93.95    \\ \cmidrule(r){1-6} \cmidrule(l){8-13} 
		\multirow{7}{*}{\begin{tabular}[c]{@{}l@{}}ResNet \\ 152\end{tabular}}                           & X                                                  & 86.14     & 84.96    & 88.57    & 86.26    &  & \multirow{7}{*}{\begin{tabular}[c]{@{}l@{}}DenseNet \\ 201\end{tabular}}                         & X                                                  & 92.98     & 95.2     & 93.37    & 95.23    \\
		& L                                                  & 86.14     & 85.92    & 88.57    & 86.63    &  &                                                                                                  & L                                                  & 93.99     & 95.36    & 95.03    & 94.87    \\
		& C                                                  & 86.14     & 87.04    & 88.74    & 85.34    &  &                                                                                                  & C                                                  & 93.82     & 95.36    & 94.53    & 93.58    \\
		& X \& L                                             & 86.47     & 85.76    & 89.56    & 86.26    &  &                                                                                                  & X \& L                                             & 94.49     & 95.04    & 94.70    & 95.05    \\
		& X \& C                                             & 86.97     & 86.24    & 88.90    & 86.81    &  &                                                                                                  & X \& C                                             & 93.65     & 94.88    & 94.20    & 94.68    \\
		& L \& C                                             & 86.31     & 85.92    & 89.40    & 86.44    &  &                                                                                                  & L \& C                                             & 94.15     & 95.68    & 94.86    & 94.32    \\
		& X \& L \& C                                        & 86.47     & 85.92    & 89.07    & 87.17    &  &                                                                                                  & X \& L \& C                                        & 94.32     & 95.52    & 95.03    & 94.50    \\ \cmidrule(r){1-6} \cmidrule(l){8-13} 
		\multirow{7}{*}{\begin{tabular}[c]{@{}l@{}}ResNet \\ 50-v2\end{tabular}}                         & X                                                  & 95.49     & 94.24    & 93.54    & 92.67    &  & \multirow{7}{*}{Xception}                                                                        & X                                                  & 92.15     & 91.04    & 92.88    & 92.85    \\
		& L                                                  & 94.65     & 95.52    & 94.20    & 94.13    &  &                                                                                                  & L                                                  & 93.65     & 92.16    & 92.05    & 94.13    \\
		& C                                                  & 94.65     & 95.52    & 93.70    & 93.04    &  &                                                                                                  & C                                                  & 90.98     & 92.00    & 91.72    & 90.47    \\
		& X \& L                                             & 95.82     & 95.84    & 94.37    & 93.77    &  &                                                                                                  & X \& L                                             & 93.65     & 92.32    & 93.04    & 93.4     \\
		& X \& C                                             & 95.82     & 95.04    & 94.03    & 92.67    &  &                                                                                                  & X \& C                                             & 92.48     & 92.32    & 91.72    & 92.12    \\
		& L \& C                                             & 94.99     & 95.68    & 94.53    & 93.40    &  &                                                                                                  & L \& C                                             & 92.98     & 92.64    & 92.38    & 93.4     \\
		& X \& L \& C                                        & 95.49     & 95.68    & 94.37    & 93.22    &  &                                                                                                  & X \& L \& C                                        & 93.15     & 92.00    & 92.38    & 93.04    \\ \cmidrule(r){1-6} \cmidrule(l){8-13} 
		\multirow{7}{*}{\begin{tabular}[c]{@{}l@{}}ResNet \\ 101-v2\end{tabular}}                        & X                                                  & 94.15     & 95.04    & 93.37    & 93.04    &  & \multirow{7}{*}{\begin{tabular}[c]{@{}l@{}}NASNet \\ Large\end{tabular}}                         & X                                                  & 94.49     & 94.08    & 93.21    & 92.12    \\
		& L                                                  & 94.82     & 96.00    & 93.87    & 95.42    &  &                                                                                                  & L                                                  & 94.82     & 94.56    & 94.2     & 93.58    \\
		& C                                                  & 93.15     & 94.88    & 94.37    & 93.58    &  &                                                                                                  & C                                                  & 94.49     & 93.44    & 94.37    & 92.49    \\
		& X \& L                                             & 94.32     & 95.2     & 93.54    & 94.68    &  &                                                                                                  & X \& L                                             & 94.65     & 95.04    & 93.54    & 93.40    \\
		& X \& C                                             & 94.49     & 95.36    & 93.54    & 93.04    &  &                                                                                                  & X \& C                                             & 93.65     & 94.40    & 94.20    & 92.30    \\
		& L \& C                                             & 94.32     & 95.36    & 93.70    & 95.05    &  &                                                                                                  & L \& C                                             & 94.65     & 94.56    & 94.20    & 93.40    \\
		& X \& L \& C                                        & 94.15     & 95.2     & 93.87    & 94.13    &  &                                                                                                  & X \& L \& C                                        & 94.49     & 94.56    & 94.03    & 93.22    \\ \cmidrule(r){1-6} \cmidrule(l){8-13} 
		\multirow{7}{*}{\begin{tabular}[c]{@{}l@{}}ResNet \\ 152-v2\end{tabular}}                        & X                                                  & 93.99     & 92.48    & 94.37    & 93.04    &  & \multirow{7}{*}{\begin{tabular}[c]{@{}l@{}}EfficientNet \\ B6\end{tabular}}                      & X                                                  & 84.97     & 85.92    & 88.24    & 88.46    \\
		& L                                                  & 94.65     & 94.08    & 94.03    & 94.13    &  &                                                                                                  & L                                                  & 86.47     & 85.6     & 88.74    & 88.64    \\
		& C                                                  & 94.49     & 93.28    & 93.21    & 93.22    &  &                                                                                                  & C                                                  & 86.47     & 85.44    & 87.91    & 89.56    \\
		& X \& L                                             & 94.49     & 94.08    & 94.03    & 93.40    &  &                                                                                                  & X \& L                                             & 85.14     & 86.08    & 88.07    & 89.01    \\
		& X \& C                                             & 94.49     & 92.80    & 93.70    & 93.40    &  &                                                                                                  & X \& C                                             & 85.47     & 85.44    & 88.24    & 89.19    \\
		& L \& C                                             & 94.65     & 93.76    & 94.03    & 94.13    &  &                                                                                                  & L \& C                                             & 86.47     & 86.08    & 88.90    & 89.74    \\
		& X \& L \& C                                        & 94.82     & 94.08    & 94.37    & 93.95    &  &                                                                                                  & X \& L \& C                                        & 85.80     & 85.76    & 88.41    & 89.19    \\
		\bottomrule
	\end{tabular*}%
\end{table*}
\clearpage
\begin{table*}[width=.99\textwidth,cols=4,pos=!h]
	\caption{Binary classification \textbf{F1-score} results on BreakHis dataset.}
	\label{table9}
	\centering
	\footnotesize
	\begin{tabular*}{\textwidth}{@{}llccccp{1cm}llcccc@{}}
		\toprule
		\multicolumn{1}{c}{\multirow{2}{*}{\begin{tabular}[c]{@{}c@{}}Feature \\ extractor\end{tabular}}} & \multicolumn{1}{c}{\multirow{2}{*}{Classifier(s)}} & \multicolumn{4}{c}{Result (F1-Score × 100)} &  & \multicolumn{1}{c}{\multirow{2}{*}{\begin{tabular}[c]{@{}c@{}}Feature \\ extractor\end{tabular}}} & \multirow{2}{*}{Classifier(s)} & \multicolumn{4}{c}{Result (F1-Score × 100)} \\ \cmidrule(lr){3-6} \cmidrule(l){10-13} 
		\multicolumn{1}{c}{}                                                                              & \multicolumn{1}{c}{}                               & 40×       & 100×      & 200×     & 400×     &  & \multicolumn{1}{c}{}                                                                              &                                & 40×       & 100×      & 200×     & 400×     \\ \cmidrule(r){1-6} \cmidrule(l){8-13} 
		\multirow{7}{*}{VGG 16}                                                                           & X                                                  & 90.23     & 87.22     & 88.45    & 84.36    &  & \multirow{7}{*}{\begin{tabular}[c]{@{}l@{}}Inception\\ v3\end{tabular}}                           & X                              & 96.13     & 95.51     & 96.21    & 95.21    \\
		& L                                                  & 90.67     & 88.17     & 89.87    & 85.18    &  &                                                                                                   & L                              & 97.06     & 96.15     & 96.10    & 96.01    \\
		& C                                                  & 90.97     & 88.89     & 90.06    & 85.50    &  &                                                                                                   & C                              & 96.60     & 95.41     & 95.48    & 95.62    \\
		& X \& L                                             & 91.48     & 88.38     & 89.47    & 85.68    &  &                                                                                                   & X \& L                         & 96.47     & 95.72     & 96.34    & 96.42    \\
		& X \& C                                             & 90.99     & 87.90     & 89.50    & 85.38    &  &                                                                                                   & X \& C                         & 96.61     & 95.60     & 96.21    & 96.16    \\
		& L \& C                                             & 90.91     & 89.22     & 89.75    & 85.93    &  &                                                                                                   & L \& C                         & 96.71     & 95.83     & 96.21    & 96.14    \\
		& X \& L \& C                                        & 91.28     & 88.45     & 89.52    & 85.42    &  &                                                                                                   & X \& L \& C                    & 96.82     & 95.94     & 96.09    & 96.42    \\ \cmidrule(r){1-6} \cmidrule(l){8-13} 
		\multirow{7}{*}{VGG 19}                                                                           & X                                                  & 90.23     & 86.50     & 88.27    & 84.47    &  & \multirow{7}{*}{\begin{tabular}[c]{@{}l@{}}Inception\\ ResNet\\ v2\end{tabular}}                  & X                              & 97.42     & 96.69     & 96.69    & 96.98    \\
		& L                                                  & 89.70     & 88.38     & 88.76    & 85.07    &  &                                                                                                   & L                              & 97.75     & 96.80     & 97.75    & 97.24    \\
		& C                                                  & 90.71     & 88.29     & 89.93    & 86.04    &  &                                                                                                   & C                              & \textbf{98.23}     & 97.04     & 97.28    & 96.98    \\
		& X \& L                                             & 90.44     & 87.46     & 89.86    & 86.41    &  &                                                                                                   & X \& L                         & 97.52     & 97.03     & 97.64    & 97.38    \\
		& X \& C                                             & 91.11     & 87.62     & 89.94    & 86.23    &  &                                                                                                   & X \& C                         & 97.76     & 97.27     & 97.41    & 96.84    \\
		& L \& C                                             & 90.14     & 88.17     & 89.35    & 86.11    &  &                                                                                                   & L \& C                         & 97.87     & 97.25     & 97.64    & 96.56    \\
		& X \& L \& C                                        & 90.52     & 87.99     & 90.24    & 86.19    &  &                                                                                                   & X \& L \& C                    & \textbf{97.76}     & \textbf{97.03}     & \textbf{97.87}    & \textbf{97.11}    \\ \cmidrule(r){1-6} \cmidrule(l){8-13} 
		\multirow{7}{*}{\begin{tabular}[c]{@{}l@{}}ResNet \\ 50\end{tabular}}                             & X                                                  & 91.70     & 92.55     & 93.16    & 93.10    &  & \multirow{7}{*}{\begin{tabular}[c]{@{}l@{}}DenseNet \\ 121\end{tabular}}                          & X                              & 95.80     & 96.58     & 95.37    & 96.42    \\
		& L                                                  & 92.98     & 92.77     & 93.35    & 92.78    &  &                                                                                                   & L                              & 95.92     & 97.05     & 96.09    & 96.41    \\
		& C                                                  & 92.54     & 92.94     & 93.62    & 92.62    &  &                                                                                                   & C                              & 95.91     & 96.23     & 95.60    & 95.88    \\
		& X \& L                                             & 92.96     & 93.24     & 93.60    & 92.87    &  &                                                                                                   & X \& L                         & 96.26     & 97.27     & 95.96    & 96.11    \\
		& X \& C                                             & 92.68     & 93.27     & 93.85    & 92.89    &  &                                                                                                   & X \& C                         & 95.79     & 96.57     & 95.70    & 96.40    \\
		& L \& C                                             & 93.21     & 92.94     & 93.49    & 93.28    &  &                                                                                                   & L \& C                         & 95.79     & 96.70     & 95.98    & 96.55    \\
		& X \& L \& C                                        & 92.77     & 93.39     & 93.73    & 93.28    &  &                                                                                                   & X \& L \& C                    & 96.25     & 96.58     & 96.45    & 96.39    \\ \cmidrule(r){1-6} \cmidrule(l){8-13} 
		\multirow{7}{*}{\begin{tabular}[c]{@{}l@{}}ResNet \\ 101\end{tabular}}                            & X                                                  & 90.49     & 90.44     & 91.04    & 90.69    &  & \multirow{7}{*}{\begin{tabular}[c]{@{}l@{}}DenseNet \\ 169\end{tabular}}                          & X                              & 96.25     & 97.16     & 95.85    & 95.97    \\
		& L                                                  & 90.64     & 91.61     & 92.34    & 91.25    &  &                                                                                                   & L                              & 96.24     & 97.61     & 95.85    & 95.28    \\
		& C                                                  & 91.14     & 91.47     & 92.92    & 92.35    &  &                                                                                                   & C                              & 95.20     & 97.27     & 96.32    & 95.86    \\
		& X \& L                                             & 90.47     & 91.30     & 92.03    & 90.98    &  &                                                                                                   & X \& L                         & 96.25     & 97.05     & 95.85    & 95.56    \\
		& X \& C                                             & 91.03     & 91.56     & 92.56    & 91.50    &  &                                                                                                   & X \& C                         & 95.79     & 96.94     & 95.84    & 95.97    \\
		& L \& C                                             & 91.26     & 91.94     & 92.65    & 91.79    &  &                                                                                                   & L \& C                         & 95.90     & 97.50     & 96.19    & 95.44    \\
		& X \& L \& C                                        & 90.80     & 91.77     & 92.54    & 91.52    &  &                                                                                                   & X \& L \& C                    & 96.03     & 97.17     & 96.20    & 95.45    \\ \cmidrule(r){1-6} \cmidrule(l){8-13} 
		\multirow{7}{*}{\begin{tabular}[c]{@{}l@{}}ResNet \\ 152\end{tabular}}                            & X                                                  & 90.54     & 89.41     & 91.83    & 89.82    &  & \multirow{7}{*}{\begin{tabular}[c]{@{}l@{}}DenseNet \\ 201\end{tabular}}                          & X                              & 95.14     & 96.58     & 95.29    & 96.42    \\
		& L                                                  & 90.43     & 90.00     & 91.93    & 90.07    &  &                                                                                                   & L                              & 95.79     & 96.71     & 96.44    & 96.14    \\
		& C                                                  & 90.62     & 90.93     & 92.02    & 89.22    &  &                                                                                                   & C                              & 95.66     & 96.69     & 96.09    & 95.17    \\
		& X \& L                                             & 90.70     & 89.97     & 92.65    & 89.82    &  &                                                                                                   & X \& L                         & 96.16     & 96.48     & 96.21    & 96.28    \\
		& X \& C                                             & 91.08     & 90.40     & 92.15    & 90.19    &  &                                                                                                   & X \& C                         & 95.58     & 96.35     & 95.86    & 96.02    \\
		& L \& C                                             & 90.66     & 90.11     & 92.52    & 90.00    &  &                                                                                                   & L \& C                         & 95.91     & 96.94     & 96.32    & 95.75    \\
		& X \& L \& C                                        & 90.74     & 90.11     & 92.25    & 90.51    &  &                                                                                                   & X \& L \& C                    & 96.04     & 96.82     & 96.45    & 95.88    \\ \cmidrule(r){1-6} \cmidrule(l){8-13} 
		\multirow{7}{*}{\begin{tabular}[c]{@{}l@{}}ResNet \\ 50-v2\end{tabular}}                          & X                                                  & 96.83     & 95.90     & 95.34    & 94.54    &  & \multirow{7}{*}{Xception}                                                                         & X                              & 94.46     & 93.55     & 94.84    & 94.66    \\
		& L                                                  & 96.24     & 96.80     & 95.84    & 95.60    &  &                                                                                                   & L                              & 95.55     & 94.31     & 94.34    & 95.60    \\
		& C                                                  & 96.27     & 96.82     & 95.50    & 94.82    &  &                                                                                                   & C                              & 93.72     & 94.16     & 94.06    & 92.99    \\
		& X \& L                                             & 97.07     & 97.03     & 95.95    & 95.33    &  &                                                                                                   & X \& L                         & 95.53     & 94.43     & 95.04    & 95.07    \\
		& X \& C                                             & 97.06     & 96.46     & 95.71    & 94.57    &  &                                                                                                   & X \& C                         & 94.74     & 94.43     & 94.03    & 94.15    \\
		& L \& C                                             & 96.48     & 96.91     & 96.08    & 95.08    &  &                                                                                                   & L \& C                         & 95.07     & 94.66     & 94.59    & 95.08    \\
		& X \& L \& C                                        & 96.84     & 96.91     & 95.95    & 94.94    &  &                                                                                                   & X \& L \& C                    & 95.18     & 94.20     & 94.56    & 94.81    \\ \cmidrule(r){1-6} \cmidrule(l){8-13} 
		\multirow{7}{*}{\begin{tabular}[c]{@{}l@{}}ResNet \\ 101-v2\end{tabular}}                         & X                                                  & 95.89     & 96.46     & 95.25    & 94.79    &  & \multirow{7}{*}{\begin{tabular}[c]{@{}l@{}}NASNet \\ Large\end{tabular}}                          & X                              & 96.15     & 95.72     & 95.11    & 94.05    \\
		& L                                                  & 96.37     & 97.15     & 95.60    & 96.56    &  &                                                                                                   & L                              & 96.37     & 96.09     & 95.81    & 95.19    \\
		& C                                                  & 95.20     & 96.35     & 95.99    & 95.20    &  &                                                                                                   & C                              & 96.14     & 95.32     & 95.97    & 94.36    \\
		& X \& L                                             & 96.02     & 96.57     & 95.34    & 96.03    &  &                                                                                                   & X \& L                         & 96.26     & 96.44     & 95.34    & 95.03    \\
		& X \& C                                             & 96.12     & 96.69     & 95.40    & 94.79    &  &                                                                                                   & X \& C                         & 95.58     & 95.99     & 95.84    & 94.21    \\
		& L \& C                                             & 96.02     & 96.69     & 95.49    & 96.29    &  &                                                                                                   & L \& C                         & 96.26     & 96.10     & 95.83    & 95.03    \\
		& X \& L \& C                                        & 95.90     & 96.58     & 95.60    & 95.62    &  &                                                                                                   & X \& L \& C                    & 96.15     & 96.11     & 95.71    & 94.91    \\ \cmidrule(r){1-6} \cmidrule(l){8-13} 
		\multirow{7}{*}{\begin{tabular}[c]{@{}l@{}}ResNet \\ 152-v2\end{tabular}}                         & X                                                  & 95.77     & 94.67     & 95.96    & 94.77    &  & \multirow{7}{*}{\begin{tabular}[c]{@{}l@{}}EfficientNet \\ B6\end{tabular}}                       & X                              & 89.73     & 90.20     & 91.70    & 91.33    \\
		& L                                                  & 96.24     & 95.81     & 95.77    & 95.60    &  &                                                                                                   & L                              & 90.70     & 90.07     & 92.07    & 91.55    \\
		& C                                                  & 96.13     & 95.23     & 95.12    & 94.92    &  &                                                                                                   & C                              & 90.83     & 90.05     & 91.50    & 92.29    \\
		& X \& L                                             & 96.12     & 95.81     & 95.75    & 95.04    &  &                                                                                                   & X \& L                         & 89.85     & 90.41     & 91.59    & 91.87    \\
		& X \& C                                             & 96.12     & 94.90     & 95.49    & 95.04    &  &                                                                                                   & X \& C                         & 90.10     & 90.03     & 91.73    & 92.02    \\
		& L \& C                                             & 96.24     & 95.58     & 95.75    & 95.60    &  &                                                                                                   & L \& C                         & 90.74     & 90.47     & 92.24    & 92.39    \\
		& X \& L \& C                                        & 96.35     & 95.80     & 95.99    & 95.47    &  &                                                                                                   & X \& L \& C                    & 90.31     & 90.25     & 91.88    & 91.99    \\
		\bottomrule
	\end{tabular*}%
\end{table*}
\clearpage

\subsection{Discussion}
Analysis of breast cancer histopathological images by utilizing image processing methods developed for the medical field has received much attention as of late. Non-automatic examination of breast cancer histopathological images is susceptible to a host of problems including observer variability, human error and an extremely tedious procedure \cite{rashmi2022breast}. A solution that alleviates these obstacles is the use of CAD systems. In this study, we proposed a new model for breast cancer histopathology images binary classification based on deep feature transfer learning. In comparing the IRv2-CXL results to other existing binary breast tumour classification methods, the proposed deep feature method attains better or comparable performance without using any image augmentation (see \autoref{table8}), signifying transfer learning and GBDTs potential in this task while demonstrating ensemble methods' power in predictive accuracy and bias/variance trade-off. With regards to improved results, we believe that a better CAD system can be developed by using IRv2-CXL as its foundation.\par
As illustrated in \autoref{table8}, IRv2-CXL achieved better results than the 16 models included in the table in three of the four magnifications factors. In the 40× magnification factor, the proposed model outperformed almost all the other models by a considerable margin, the only close ones being Sharma and Kumar \cite{sharma2022xception} with 96.25\% and Kumar et al. \cite{kumar2020deep} with 94.11\%. IRv2-CXL attained a 95.84\% accuracy in the 100× magnification factor, only outdone by the 96.25\% achieved by Sharma and Kumar \cite{sharma2022xception} with a 0.41\% margin. The 200× magnification factor led to the best individual magnification result of the proposed model. In this magnification, IRv2-CXL and the Kumar et al. \cite{kumar2020deep} model surpassed the other models with at best, a 1.27\% and worst, an 18.91\%  accuracy gap by achieving the same 97.01\% accuracy. In the 400× magnification factor, the proposed model secured the best result by a 2.04\% margin with a 96.15\% accuracy score; the second-best model being Sharma and Kumar's \cite{sharma2022xception}. On average IRv2-CXL model attained the best accuracy, 96.46\%,  followed by Sharma and Kumar's \cite{sharma2022xception} model with a 95.58\% average accuracy score. \par
We decided to use GBDT based methods as classifier because of their high prediction accuracy, and adaptability to non-linear characteristics and good training effect \cite{xu2021intelligent, wang2019multiple}. XGBoost, CatBoost and LightGBM were chosen as classifier candidates. XGBoost was chosen because of its new regularization method that resists overfitting, therefore making the model more robust \cite{xu2021intelligent,fatahi2022modeling}. CatBoosts improved generalization as well as its ability to capture high-order dependencies makes it a viable candidate as well \cite{zhang2021fault}. Last but not least LightGBM is an attractive method to use as a classifier because of its capability in handling large-scale data. In addition, CatBoost's leaf-wise growth approach makes it more accurate than other level-wise GBDT methods \cite{xu2021intelligent, zhang2019predictive}. As all three of our classifier candidates performed relatively well on this task, we decided to incorporate a soft majority voting method in the proposed method to enhance its predictive performance and improve the model’s robustness by lowering its variance. We found that IRv2-CXL, a model with Inception-ResNet-v2 as its feature extractor and soft majority voting ensemble of XGBoost, CatBoost and LightGBM achieves better magnification-wise average accuracy than other models we experimented on. \par
IRv2-CXL has several advantages in comparison to other existing models. By using a pre-trained model as feature extractor, its training time is much less than those models that attempt to extract features from images themselves. The use of transfer learning also reduces the proposed model's need for huge amounts of training data which is almost always in short supply in the medical field. Incorporating an ensemble of GBDT based algorithms in the IRv2-CXL classifier not only improves its accuracy in comparison to other models but also draws attention to the fact that not enough scrutiny has been afforded to choosing classifiers in previous studies and most of the researchers tackling this task have focused on developing better feature extractors \cite{chan2016automatic}. The use of soft majority voting provides IRv2-CXL with a lower variance relative to other models. As the proposed model performs well with all magnifications present in BreakHis and has low accuracy variance across all of them, a CAD system based on it would have an extremely low dependency on the magnification factor. While the proposed model has many advantages, it has weaknesses as well. The first weakness stems from a lack of feature extractor fine-tuning. Inception-ResNet-v2’s large image data belongs mostly to the general domain, such as ear, corn and vase, however, this task’s images are breast cancer histopathology images with non-identical visual appearances which may make IRv2-CXL's feature extractor biased with the source data and less generalized on the target data \cite{shalbaf2021automated}. The second weakness arises from the dataset we used to develop this model and the blatant imbalance in its class distributions. while our model performs well on the F1 score (\autoref{table6}) and ROC curve (\autoref{table5}) as well as accuracy and its high accuracy is not reliant on this imbalance, we believe a more balanced dataset would produce even better results. \par
In the future, we aim to resolve the proposed model's lack of generalization on the target dataset through fine-tuning and addressing the imbalanced dataset issue with a more comprehensive preprocessing including an image augmentation that makes the model more balanced. Further improvements can be achieved by incorporating pathologist expertise and the grad-cam images we have already produced. These two can help us determine the nonconformities between the areas that help the expert and IRv2-CXL decide on tumour classification, then we can construct an attention-based model targeting those areas most helpful to correct classification. \par

\begin{table*}[width=.99\textwidth,cols=4,pos=p]
	\caption{Comparison of IRv2-CXL with other models.}
	\label{table8}
	\footnotesize
	\begin{tabular*}{\tblwidth}{@{} CCCCCC @{}}
		\toprule
		\multirow{2}{*}{Study (Year)} & \multicolumn{5}{c}{Results (Accuracy \%)} \\ \cmidrule(l){2-6} 
												   		   & 40×    & 100×   & 200×  & 400×  & Average \\
		
		\midrule
		\multicolumn{1}{l|}{\begin{tabular}[c]{@{}c@{}}Spanhol et al. \cite{spanhol2017deep} ~~ (2017) \end{tabular}}       		     & 84.60  			& 84.80 			& 84.20 			& 81.60 			& 83.80   \\
		\multicolumn{1}{l|}{\begin{tabular}[c]{@{}c@{}}Song et al. \cite{song2017adapting} ~~ (2017) \end{tabular}}           		     & 87.00  			& 86.20  			& 85.20 			& 82.90 			& 85.32   \\
		\multicolumn{1}{l|}{\begin{tabular}[c]{@{}c@{}}Song et al. \cite{song2017supervised}  ~~ (2017) \end{tabular}}       		     & 87.70  			& 87.60  			& 86.50 			& 83.90 			& 86.42   \\
		\multicolumn{1}{l|}{\begin{tabular}[c]{@{}c@{}}Alirezazadeh et al. \cite{alirezazadeh2018representation} ~~ (2018) \end{tabular}}& 89.10   			& 87.30     		& 91.00    			& 86.60    			& 88.50   \\
		\multicolumn{1}{l|}{\begin{tabular}[c]{@{}c@{}}Sanchez et al. \cite{sanchez2018classification} ~~ (2018) \end{tabular}} 		 & 85.90  			& 80.40  			& 78.10 			& 71.10 			& 78.87   \\
		\multicolumn{1}{l|}{\begin{tabular}[c]{@{}c@{}}Benhammou et al. \cite{benhammou2018first} ~~ (2018) \end{tabular}}     			 & 85.50  			& 83.20  			& 85.40 			& 80.30 			& 83.60   \\
		\multicolumn{1}{l|}{\begin{tabular}[c]{@{}c@{}}Du et al. \cite{du2018breast} ~~ (2018) \end{tabular}}                  			 & 90.69  			& 90.46  			& 90.64 			& 90.96 			& 90.68   \\
		\multicolumn{1}{l|}{\begin{tabular}[c]{@{}c@{}}Badejo et al. \cite{badejo2018medical} ~~ (2018) \end{tabular}}        		     & 91.10  			& 90.70  			& 86.20 			& 84.30 			& 88.07   \\
		\multicolumn{1}{l|}{\begin{tabular}[c]{@{}c@{}}Nahid et al. \cite{nahid2018histopathological} ~~ (2018) \end{tabular}}  		 & 88.70  			& 85.30  			& 88.60 			& 88.40 			& 87.75   \\
		\multicolumn{1}{l|}{\begin{tabular}[c]{@{}c@{}}Kumar and Rao \cite{kumar2018breast} ~~ (2018) \end{tabular}}          		     & 82.00  			& 86.20  			& 84.60 			& 84.00 			& 84.20   \\
		\multicolumn{1}{l|}{\begin{tabular}[c]{@{}c@{}}Sudharshan et al. \cite{sudharshan2019multiple} ~~ (2019) \end{tabular}} 		 & 87.80  			& 85.60  			& 80.80 			& 82.90 			& 84.27   \\
		\multicolumn{1}{l|}{\begin{tabular}[c]{@{}c@{}}Kumar et al. \cite{kumar2020deep} ~~ (2020) \end{tabular}}     		             & 94.11  			& 95.12  			& \textbf{97.01}	& 93.40 			& 94.91   \\
		\multicolumn{1}{l|}{\begin{tabular}[c]{@{}c@{}}Li et al. \cite{li2021classification} ~~ (2021) \end{tabular}}          		     & 87.85 			& 86.68  			& 87.75 			& 85.30 			& 86.89   \\
		\multicolumn{1}{l|}{\begin{tabular}[c]{@{}c@{}}Sharma and Kumar \cite{sharma2022xception} ~~ (2022) \end{tabular}}     		     & 96.25 			& \textbf{96.25}    & 95.74 			& 94.11 			& 95.58   \\
		\multicolumn{1}{l|}{\begin{tabular}[c]{@{}c@{}}Joseph et al. \cite{joseph2022improved} ~~ (2022) \end{tabular}}          	     & 90.87  			& 89.57  			& 91.58 			& 88.67 			& 90.17   \\
		\multicolumn{1}{l|}{\begin{tabular}[c]{@{}c@{}}Zerouaoui and Idri \cite{zerouaoui2022deep} ~~ (2022) \end{tabular}}     		 & 92.61  			& 92.00  			& 93.93 			& 91.73 			& 92.56   \\
		\multicolumn{1}{l|}{} &&&&& \\
		\multicolumn{1}{l|}{\begin{tabular}[c]{@{}c@{}}IRv2-CXL ~~ (2022) \end{tabular}}                                         & \textbf{96.82}   & 95.84  			& \textbf{97.01}	& \textbf{96.15}	& \textbf{96.46} \\
		
		\bottomrule
	\end{tabular*}
\end{table*}

\newpage
\section{Conclusion}
In this paper, we tackled the issue of breast tumour type (benign or malignant) estimation by processing histopathological images obtained from the lab. This tumour type assessment remains an unsolved problem in early cancer detection that by resolving it with the help of an automatic classification system we can contribute greatly to the medical community and ordinary people, saving them money, time and above all lives. \par
IRv2-CXL, a novel binary classification method for CAD systems is introduced in this work. Most CAD systems consist of three phases, we chose to concentrate on the classification phase. This approach implements deep features extracted from a pre-trained network and a classifier to distinguish between benign and malignant tumours. 16 different pre-trained networks were tested with seven classifiers (three single and four ensembles). The best model was found to be an Inception-ResNet-v2 network and a soft majority voting ensemble of gradients boosted decision trees (i.e., CatBoost,...) named IRv2-CXL. IRv2-CXL was trained on and then validated using the BreakHis dataset. IRv2-CXL achieved an accuracy of 96.83\%, 95.84\%, 97.02\%, and 96.16\% on the 40×, 100×, 200×, and 400× magnification factors, respectively. Furthermore, a class activation mapping was also implemented to identify important regions in the image that help IRv2-CXL classify tumours to make sure the proposed model is not concentrating on any unrelated parts. \par
We believe that further improvements to this model or the ideas it is based on are entirely possible and within reach. We hope that this study will attract others to explore the ideas presented and build upon our and other existing research findings to introduce even better methods with more outstanding results.  \par

\section{Data and Code Availability}
The BreakHis dataset (version 1) is publicly available \href{https://web.inf.ufpr.br/vri/databases/breast-cancer-histopathological-database-breakhis/}{here}.
In addition, the source code of the proposed model required to reproduce the predictions and results is available at the public \href{https://github.com/mohammadAbbasniya/BreastCancer-Detection}{Github repository}.

\section{Declaration of Competing Interest}
The authors express that they have no competing interest affecting the reported work in this article.

\bibliographystyle{unsrt}

\end{document}